\documentclass[prd,reprint,longbibliography,nofootinbib,superscriptaddress]{revtex4-2}

\usepackage[USenglish]{babel}
\usepackage[T1]{fontenc}
\usepackage{amsmath}
\usepackage{mathtools}
\usepackage{amssymb}
\usepackage{dsfont}
\usepackage{graphicx}
\usepackage{physics}
\usepackage[acronym]{glossaries}
\usepackage{siunitx}
\usepackage[svgnames]{xcolor}
\usepackage[hidelinks]{hyperref}
\hypersetup{
	colorlinks = true, 
	urlcolor = DarkBlue, 
	linkcolor = DarkBlue, 
	citecolor = DarkBlue  
}
\usepackage{cleveref}
\creflabelformat{equation}{#2\textup{#1}#3}

\begin{document}
	
\title{Time-delay interferometry with onboard optical delays}
	
\author{Jan Niklas Reinhardt}
\email{janniklas.reinhardt@aei.mpg.de}
\affiliation{Max-Planck-Institut für Gravitationsphysik (Albert-Einstein-Institut),\\ Callinstraße 38, 30167 Hannover, Germany}
\affiliation{Leibniz Universität Hannover, Welfengarten 1, 30167 Hannover, Germany}

\author{Philipp Euringer}
\affiliation{Airbus Space Systems, Airbus Defence and Space GmbH,\\Claude-Dornier-Straße, 88090 Immenstaad am Bodensee, Germany}

\author{Olaf Hartwig}
\affiliation{Max-Planck-Institut für Gravitationsphysik (Albert-Einstein-Institut),\\ Callinstraße 38, 30167 Hannover, Germany}
\affiliation{Leibniz Universität Hannover, Welfengarten 1, 30167 Hannover, Germany}
\affiliation{Max-Planck-Institut für Gravitationsphysik (Albert-Einstein-Institut),\\ Am Mühlenberg 1, 14476 Potsdam, Germany}

\author{Gerald Hechenblaikner}
\affiliation{Airbus Space Systems, Airbus Defence and Space GmbH,\\Claude-Dornier-Straße, 88090 Immenstaad am Bodensee, Germany}

\author{Gerhard Heinzel}
\affiliation{Max-Planck-Institut für Gravitationsphysik (Albert-Einstein-Institut),\\ Callinstraße 38, 30167 Hannover, Germany}
\affiliation{Leibniz Universität Hannover, Welfengarten 1, 30167 Hannover, Germany}

\author{Kohei Yamamoto}
\affiliation{Max-Planck-Institut für Gravitationsphysik (Albert-Einstein-Institut),\\ Callinstraße 38, 30167 Hannover, Germany}
\affiliation{Leibniz Universität Hannover, Welfengarten 1, 30167 Hannover, Germany}


\begin{abstract}
    Time-delay interferometry (TDI) is a data processing technique for space-based gravitational-wave detectors to create laser-noise-free equal-optical-path-length interferometers virtually on the ground. It relies on the interspacecraft signal propagation delays, which are delivered by intersatellite ranging monitors. Also, onboard signal propagation and processing delays have a non-negligible impact on the TDI combinations. However, these onboard delays were only partially considered in previous TDI-related research; onboard optical path lengths have been neglected. In this paper, we study onboard optical path lengths in TDI. We derive analytical models for their coupling to the second-generation TDI Michelson combinations and verify these models numerically. Furthermore, we derive a compensation scheme for onboard optical path lengths in TDI and validate its performance via numerical simulations.
\end{abstract}

\maketitle

\section{Introduction}\label{sec:Introduction}
The laser interferometer space antenna (LISA) is a future space-based gravitational-wave detector with a sensitive detection bandwidth between \SI{0.1}{\milli\hertz} and \SI{1}{\hertz} \cite{Colpi:2024RedBook}. It consists of three spacecraft (SC) on heliocentric orbits spanning a triangular configuration with an arm length of about 2.5 \si{\giga\m}. Gravitational waves cause picometer arm-length variations in the LISA constellation, which are detected via laser interferometry in interspacecraft interferometers (ISIs).\par 
Each SC contains two lasers with a nominal wavelength of \SI{1064}{\nano\m}. They are sent to the other two SC to set up six laser links between the three LISA satellites. The Doppler shifts due to the relative SC motion necessitate heterodyne interferometry between received and local lasers in the ISI. The corresponding beatnotes are detected with quadrant photoreceivers (QPRs),\footnote{The six lasers are offset frequency locked to each other according to a predetermined frequency plan \cite{Heinzel:2024FrequencyPlanning}. This constrains the beatnote frequencies in the sensitive QPR detection bandwidth (\SIrange{5}{25}{\mega\hertz}), thus counteracting time-varying Doppler shifts.} followed by phase extraction using digital phasemeters \cite{Gerberding:Phasemeter}. The picometer arm length variations due to gravitational waves manifest as microcycle phase fluctuations in the ISI beatnotes, which defines the target sensitivity.\par
However, laser frequency noise exceeds this target sensitivity by more than eight orders of magnitude. This led to the development of time-delay interferometry (TDI), which is an on-ground data processing technique to mitigate laser frequency noise \cite{Armstrong:TDI,Tinto:TDIforLISA,Tinto:2020TDIReview}. TDI relies on measurements of the interspacecraft signal propagation delays (interspacecraft ranging) \cite{Heinzel:Ranging,Sutton:Ranging} to compose equal-optical-path-length interferometers from the LISA interferometric measurements. These TDI combinations naturally cancel laser frequency noise.\par
Each SC houses two free-falling test masses \cite{Armano:LPF-TestMasses,Armano:BeyondLPF}.\footnote{To be precise, the test masses are free-falling only along the respective intersatellite axes. In the other directions, electrostatic forces are applied to keep them stable.} They are decoupled from the optical benches (OBs) and, thus, from the ISIs, which measure distance variations between local and distant OBs. From the perspective of TDI, the test masses can be considered as the start and end points of the intersatellite laser links. The above-mentioned TDI combinations are defined between the test masses, which act as free-falling mirrors in the virtual interferometers. Hence, TDI requires measurements of the interspacecraft-test-mass-to-test-mass separations as building blocks for the virtual equal-optical-path-length interferometers. This necessitates further interferometers to measure the OB motion with respect to the free-falling test masses: The test-mass interferometer (TMI) and the reference interferometer (RFI). These interferometers utilize the two lasers onboard the same spacecraft, whose beams are exchanged between the two OBs using an optical fiber. Within the framework of TDI, we combine ISI, TMI, and RFI beatnotes to set up the measurements of the interspacecraft-test-mass-to-test-mass separations \cite{Otto:Thesis}. Usually, this step is referred to as the removal of the optical bench jitter.\par
\begin{figure*}
	\begin{center}
		\includegraphics[width=1.0\textwidth]{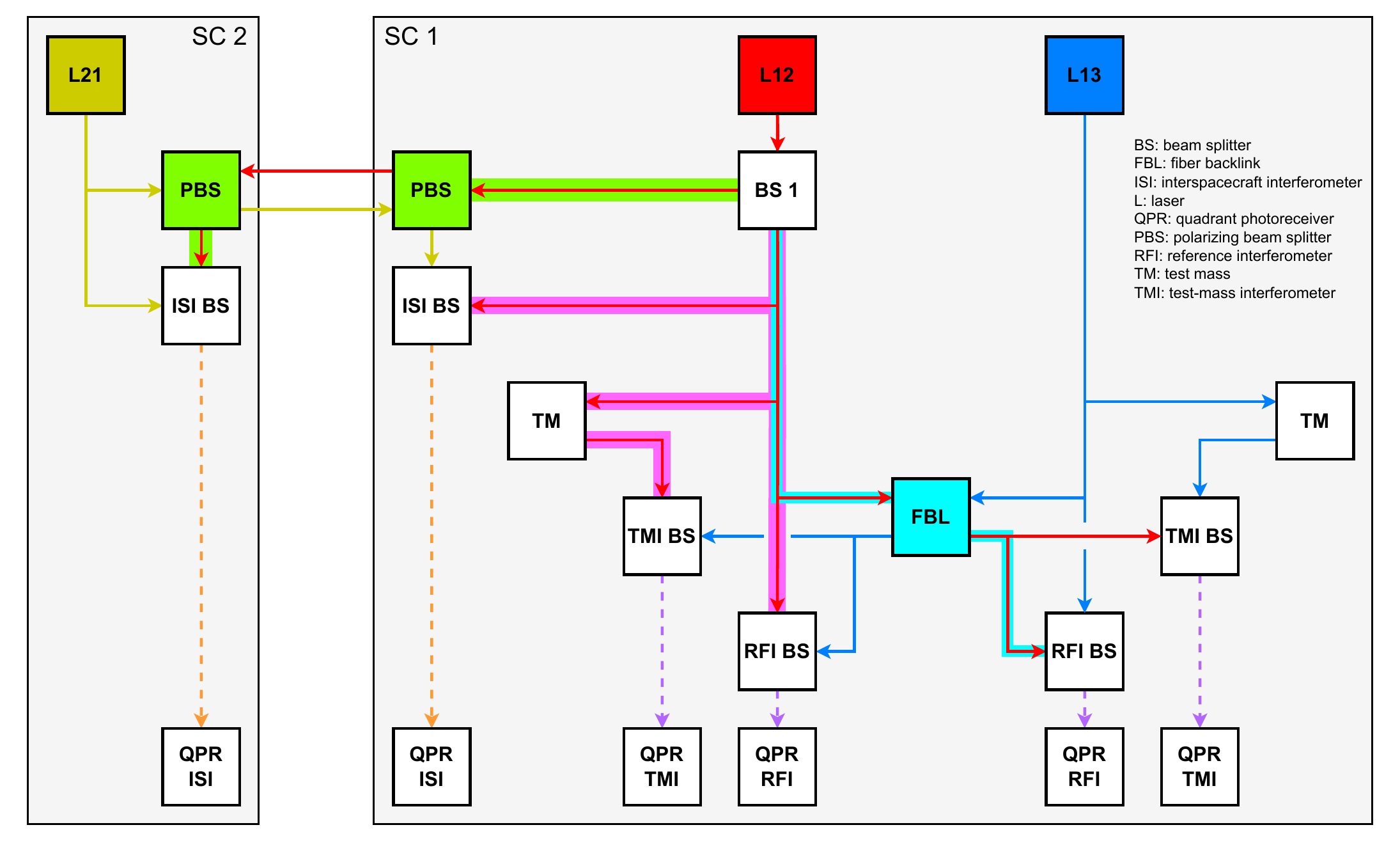}
	\end{center}
	\caption{We trace laser 12 (red arrows) to the local and to the distant ISI, where it interferes with the distant laser 21 (yellow arrows) to form beatnotes (orange dashed arrows). We further trace laser 12 to the local TMI and RFI and to the adjacent TMI and RFI, where it interferes with the adjacent laser 13 (blue arrows) to form beatnotes (purple dashed arrows). The reference point is the BS dividing outgoing and local beams (BS 1 according to the notation in \cite{Brzozowski:LISA-OB}). The OOPLs from BS 1 to the combining BSs at the local, adjacent, and distant interferometers are highlighted pink, light blue, and green, respectively.}\label{fig:OnboardOpticalPathLengths}
\end{figure*}
Apart from interspacecraft signal propagation delays, also delays due to onboard signal propagation and processing emerge in the TDI combinations. However, these onboard delays were mostly neglected or only partially considered in previous TDI-related research. Onboard delays can be grouped into two categories: (1) Onboard delays that occur after the combining beam splitters (BSs) at the different interferometers are common to both interfering beams, e.g., electronic delays in the QPRs and signal processing delays in the phasemeter. A detailed investigation of common onboard delays can be found in \cite{Euringer:FrontendAndModulationDelays,Yamamoto:2024Hexagon}. (2) Onboard delays before the combining BSs differ between both interfering beams. These are onboard optical path lengths (OOPLs) between the laser sources and the combining BSs (see \cref{fig:OnboardOpticalPathLengths}). \cite{Reinhardt:RangingSensorFusion} suggests a compensation method for OOPLs in the ISI but neglects TMI and RFI. However, the OOPLs in TMI and RFI are expected to dominate due to the potentially several meters long fiber backlink. If uncompensated, they cause residual laser noise in the TDI combinations. While previous research established models for the coupling of ISI OOPLs in TDI, where they act as ranging biases \cite{Staab:LaserNoiseResiduals,Staab:Thesis}, we lack such models for the coupling of TMI and RFI OOPLs.\par
This paper studies the TDI coupling of OOPLs in all interferometers. In \cref{sec:BeatNotesWithOOPLs}, we introduce delay and advancement operators for OOPLs. This allows us to express the LISA beatnotes, including OOPLs. In \cref{sec:TDIwithOOPLs}, we derive a compensation scheme for OOPLs, which includes the OOPL delay and advancement operators in the TDI processing steps. We derive analytical models for the TDI coupling of OOPLs in \cref{sec:LaserNoiseResidualsDueToOOPLs}. We numerically implement the OOPL compensation scheme and demonstrate its performance in \cref{sec:Results}, where we further compare the numerical results with the analytical models. We conclude in \cref{sec:Conclusion}.

\section{LISA beatnotes with onboard optical path lengths}\label{sec:BeatNotesWithOOPLs}

\subsection{Brief summary of the LISA payload}
\begin{figure}
	\begin{center}
		\includegraphics[width=0.5\textwidth]{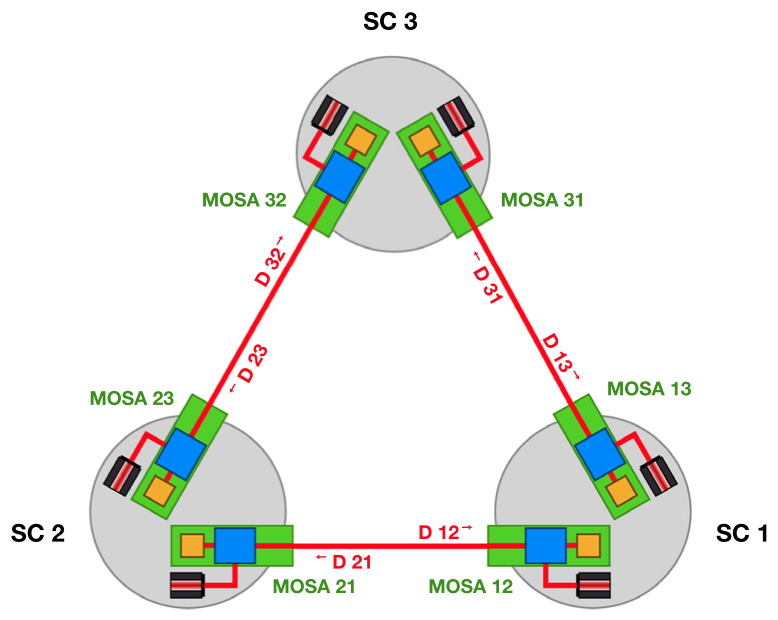}
	\end{center}
	\caption{LISA labeling conventions (from \cite{Bayle:TDI-in-Frequency}). The SC are labeled clockwise. The MOSAs and associated building blocks (lasers, interferometers, etc.) are labeled with 2 indices: The first one denotes the SC they are located on, the second one the SC they are facing. Left-handed MOSAs are labeled 12, 23, 31. Right-handed MOSAs are labeled 13, 32, 21. Delays are labeled according to the MOSA in which they can be measured, e.g., the delay of the beam received by SC 1 from SC 2 is labeled $\textbf{D}_{12}$.}\label{fig:LisaLabelingConventions}
\end{figure}
Each SC contains two movable optical subassemblies (MOSAs), which are oriented toward the other two SC of the constellation (the labeling conventions are summarized in \cref{fig:LisaLabelingConventions}). Each MOSA has a laser, which is fiber-fed to an optical bench (OB) made of Zerodur. From there, the laser is transmitted to the distant SC via a telescope and to the adjacent MOSA via an optical fiber (the backlink). On the OB itself, the laser serves as a local oscillator in three heterodyne interferometers: The interspacecraft interferometer (ISI) interferes the local beam with the beam received from the distant SC; the test-mass interferometer (TMI) and the reference interferometer (RFI) interfere the local beam with the beam received from the adjacent MOSA. Before interference in the TMI, the local beam is reflected off a free-falling cubic gold-platinum test mass along the intersatellite axis.

\subsection{Delay operators}
We neglect clock desynchronizations \cite{Reinhardt2024:ClockSynchronization} and express all quantities in terms of a virtual LISA constellation time $\tau$. Following the notation of \cite{Bayle:LISA-Simulation}, we write the frequency of laser $ij$ in terms of its offset $O_{ij}$ from the nominal laser frequency $\nu_0$ and laser frequency noise $\dot{p}_{ij}$:
\begin{align}\label{eq:LaserFrequency}
	\nu_{ij}(\tau) = \nu_0 + O_{ij}(\tau) + \dot{p}_{ij}(\tau).
\end{align}
This frequency is defined at the laser source. To model the LISA beatnotes, we must compare the beam frequencies at the combining beam splitters (BSs) of the different interferometers. This requires the concept of delay operators.\par
The interspacecraft delay operator $\textbf{D}_{ij}$ delays the time argument of the beam phase $\Phi_{ji}$ by the intersatellite signal propagation time\footnote{To decouple interspacecraft signal propagation delays from onboard delays, we define $d_{ij}$ between the polarizing beam splitters (PBSs) in front of the telescopes on the receiving and emitting SC (see \cref{fig:OnboardOpticalPathLengths}) \cite{Reinhardt:RangingSensorFusion}.} to SC $i$ from SC $j$, denoted $d_{ij}$:
\begin{align}\label{eq:InterSpacecraftDelayOperator}
	\textbf{D}_{ij}\:\Phi_{ji}(\tau) &= \Phi_{ji}\left(\tau - d_{ij}(\tau)\right).
\end{align}
Here, we express the LISA beatnotes in frequency, i.e., we must take the time derivative of \cref{eq:InterSpacecraftDelayOperator}:
\begin{align}\label{eq:InterSpacecraftDopplerDelayOperator}
	\dot{\textbf{D}}_{ij}\:\nu_{ji}(\tau) \vcentcolon= \left(1 - \dot{d}_{ij}(\tau)\right)\: \nu_{ji}\left(\tau - d_{ij}(\tau)\right),
\end{align}
where $\dot{\textbf{D}}_{ij}$ denotes the Doppler-delay operator \cite{Bayle:TDI-in-Frequency}.\par
In addition to interspacecraft signal propagation delays, we must consider delays due to onboard signal propagation and processing. We briefly revisit our categorization of onboard delays from \cref{sec:Introduction}: (1) Onboard delays occurring after the combining BSs are common to both interfering beams. We here neglect common onboard delays. They can be compensated by time-shifting the beatnotes in an initial data treatment \cite{Euringer:FrontendAndModulationDelays}. (2) Onboard delays occurring before the combining BSs differ between both interfering beams. These are delays due to onboard optical path lengths (OOPLs) between the laser sources and the combining BSs (see \cref{fig:OnboardOpticalPathLengths}). They are the subject of this paper. Below, we neglect the conversion between optical path lengths and the associated optical delays and use these terms interchangeable.\par
To express the LISA beatnotes, including OOPLs, we introduce the onboard delay operator for OOPLs:
\begin{align}\label{eq:OnboardOpticalDelayOperator}
    \dot{\textbf{D}}_{\text{ifo}}^{\text{beam}}\:\nu_{ij}(\tau) &= \left(1 - \dot{d}_{\text{ifo}}^{\text{beam}}\right)\nu_{ij}\left(\tau - d_{\text{ifo}}^{\text{beam}}\right)\nonumber\\
    &\approx \nu_{ij}\left(\tau - d_{\text{ifo}}^{\text{beam}}\right) =\vcentcolon \textbf{D}_{\text{ifo}}^{\text{beam}}\:\nu_{ij}(\tau).
\end{align}
\textbf{ifo} is a placeholder for the target interferometer. Thus, it takes on the symbols \textbf{isi}, \textbf{tmi}, and \textbf{rfi}. \textbf{beam} is a placeholder for the particular beam. It distinguishes between local, adjacent, and distant beams denoted by \textbf{loc}, \textbf{adj}, and \textbf{dist}. We consider OOPLs constant at the scales applicable for laser noise suppression and neglect the associated Doppler terms.\footnote{In principle, longitudinal OB jitter can be regarded as a time-varying OOPL. However, its magnitude in the order of \SIrange[]{1}{10}{\nano \m \hertz \tothe{-0.5}} \cite{Inchauspe:2022SCjitter} is completely negligible in the context of laser frequency noise suppression.}\par
For a constant delay operator $\textbf{D}_x$ we define the associated advancement operator $\textbf{A}_x$, which acts as its inverse:
\begin{align}\label{eq:AdvancementOperator}
	\textbf{A}_x\: \nu_{ij}(\tau) \vcentcolon=&\: \nu_{ij}\left(\tau + x\right), \\
	\textbf{A}_x\:\textbf{D}_x\:\nu_{ij}(\tau) =&\: \nu_{ij}\left(\tau- x + x\right) = \nu_{ij}(\tau),
\end{align}
where $x$ is a placeholder for the concrete delay.

\subsection{LISA beatnotes with onboard optical delays}
Each laser enters six interferometers: the ISI, TMI, and RFI on the local MOSA, the TMI and RFI on the adjacent MOSA, and the ISI on the distant MOSA. The OOPL delay operator (see \cref{eq:OnboardOpticalDelayOperator}) allows us to write the LISA beatnotes, including the OOPLs between the laser sources and the combining BSs at these interferometers (see \cref{fig:OnboardOpticalPathLengths}):
\begin{align}
    \text{isi}_{ij} &= \textbf{D}_{\text{isi}}^{\text{in}}\:\dot{\textbf{D}}_{ij}\:\textbf{D}_{\text{isi}}^{\text{out}}\:\nu_{ji} - \textbf{D}_{\text{isi}}^{\text{loc}}\:\nu_{ij}+\dot{N}_{ij}^{\text{isi}},\label{eq:TotalIsiBeatnote}\\
    \text{tmi}_{ij} &= \textbf{D}_{\text{tmi}}^{\text{adj}}\:\nu_{ik} - \textbf{D}_{\text{tmi}}^{\text{loc}}\:\nu_{ij} + \dot{N}_{ij}^{\text{tmi}},\label{eq:TotalTmiBeatnote}\\
    \text{rfi}_{ij} &= \textbf{D}_{\text{rfi}}^{\text{adj}}\:\nu_{ik} - \textbf{D}_{\text{rfi}}^{\text{loc}}\:\nu_{ij} + \dot{N}_{ij}^{\text{rfi}},\label{eq:TotalRfiBeatnote}
\end{align}
$\textbf{D}_{\text{isi}}^{\text{out}}$ and $\textbf{D}_{\text{isi}}^{\text{in}}$ denote the OOPLs of the distant beam on the distant and local SC, respectively. Laser frequency noise $\dot{p}_{ij}$ is included in the laser frequency (see \cref{eq:LaserFrequency}). All other noises are summarized in $\dot{N}^{\text{ifo}}_{ij}$.\par
Like the individual laser frequencies, the LISA beatnotes can be decomposed into large offsets and small fluctuations. We want to study laser frequency noise cancellation in the presence of OOPLs, so we focus on the latter and write \cref{eq:TotalIsiBeatnote,eq:TotalTmiBeatnote,eq:TotalRfiBeatnote} as
\begin{align}
    \text{isi}_{ij} &= \textbf{D}_{\text{isi}}^{\text{in}}\:\dot{\textbf{D}}_{ij}\:\textbf{D}_{\text{isi}}^{\text{out}}\:\dot{p}_{ji} - \textbf{D}_{\text{isi}}^{\text{loc}}\:\dot{p}_{ij},\label{eq:IsiBeatnoteFluctuations}\\
    \text{tmi}_{ij} &= \textbf{D}_{\text{tmi}}^{\text{adj}}\:\dot{p}_{ik} - \textbf{D}_{\text{tmi}}^{\text{loc}}\:\dot{p}_{ij},\label{eq:TmiBeatnoteFluctuations}\\
    \text{rfi}_{ij} &= \textbf{D}_{\text{rfi}}^{\text{adj}}\:\dot{p}_{ik} - \textbf{D}_{\text{rfi}}^{\text{loc}}\:\dot{p}_{ij},\label{eq:RfiBeatnoteFluctuations}
\end{align}
where we dropped the noise terms $\dot{N}^{\text{ifo}}_{ij}$ for the ease of notation. We can commute $\dot{\textbf{D}}_{ij}$ and $\textbf{D}_{\text{isi}}^{\text{in}}$, since
\begin{align}
    &\:\textbf{D}_{\text{isi}}^{\text{in}}\:\dot{\textbf{D}}_{ij}\:\dot{p}(\tau)\nonumber\\
    =&\:\left(1 - \dot{d}_{ij}(\tau-d_{\text{isi}}^{\text{in}})\right)\dot{p}\left(\tau - d_{\text{isi}}^{\text{in}} - d_{ij}\left(\tau - d_{\text{isi}}^{\text{in}}\right)\right)\nonumber\\
    \approx&\:\left(1 - \dot{d}_{ij} + \ddot{d}_{ij}\cdot d_{\text{isi}}^{\text{in}}\right)\dot{p}\left(\tau - d_{\text{isi}}^{\text{in}} - d_{ij} + \dot{d}_{ij} \cdot d_{\text{isi}}^{\text{in}} \right)\nonumber\\
    \approx&\:\left(1 - \dot{d}_{ij}(\tau)\right)\dot{p}\left(\tau - d_{\text{isi}}^{\text{in}} - d_{ij}(\tau) \right)\nonumber\\
    =&\:\dot{\textbf{D}}_{ij}\:\textbf{D}_{\text{isi}}^{\text{in}}\:\dot{p}(\tau),
\end{align}
where we applied the following approximations (considering ESA orbits \cite{Bayle:LisaOrbits}):
\begin{align}
    c\cdot \dot{d}_{ij} \cdot d_{\text{isi}}^{\text{in}} &\lessapprox \SI{10}{\m\s\tothe{-1}} \cdot \SI{10}{\nano\s} = \SI{0.1}{\micro\m},\\
c\cdot \ddot{d}_{ij} \cdot d_{\text{isi}}^{\text{in}} &\lessapprox \SI{10}{\micro\m\s\tothe{-2}} \cdot \SI{10}{\nano\s} = \SI{0.1}{\pico\m\s\tothe{-1}}.
\end{align}
These terms are negligible considering the achievable millimeter accuracy for intersatelite ranging \cite{Reinhardt:RangingSensorFusion}. Hence, we can write
\begin{align}
    \text{isi}_{ij} =&\:\dot{\textbf{D}}_{ij}\:\textbf{D}_{\text{isi}}^{\text{dist}}\:\dot{p}_{ji} - \textbf{D}_{\text{isi}}^{\text{loc}}\:\dot{p}_{ij},\\
    \textbf{D}_{\text{isi}}^{\text{dist}} \vcentcolon=&\:\textbf{D}_{\text{isi}}^{\text{in}}\:\textbf{D}_{\text{isi}}^{\text{out}}.
\end{align}\par
In summary, we need to consider six OOPLs: $\textbf{D}_{\text{isi}}^{\text{loc}}$, $\textbf{D}_{\text{isi}}^{\text{dist}}$, $\textbf{D}_{\text{tmi}}^{\text{loc}}$, $\textbf{D}_{\text{tmi}}^{\text{adj}}$, $\textbf{D}_{\text{rfi}}^{\text{loc}}$, and $\textbf{D}_{\text{rfi}}^{\text{adj}}$. Their current design values can be roughly estimated from \cite{Brzozowski:LISA-OB}; we list these estimates in the first column of \cref{tab:OnboardOpticalPathLengths}. We expect manufacturing asymmetries in the order of \SIrange[]{10}{100}{\micro\m}. However, their impact is negligible across the LISA band (see \cref{app:ManufacturingAsymmetries}). Hence, OOPLs are well determined by their design values. Consequently, we neglect manufacturing asymmetries in the notation, i.e., we do not specify MOSA indices for OOPL delay and advancement operators.
\begin{table}[]
    \centering
    \begin{tabular}{|l|r|r|}
        \hline
        & Current OB Design & Matched OB Design\\
        \hline
        $d_{\text{isi}}^{\text{loc}}$ & \SI{0.31}{\m} & \SI{0.31}{\m}\\
        $d_{\text{isi}}^{\text{dist}}$ & \SI{0.59}{\m} & \SI{0.59}{\m}\\
        $d_{\text{tmi}}^{\text{loc}}$ & \SI{0.41}{\m} & \SI{0.41}{\m}\\
        $d_{\text{tmi}}^{\text{adj}}$ & \SI{10.71}{\m} & \SI{10.71}{\m}\\
        $d_{\text{rfi}}^{\text{loc}}$ & \SI{0.36}{\m} & \SI{0.36}{\m}\\
        $d_{\text{rfi}}^{\text{adj}}$ & \SI{10.64}{\m} & \SI{10.66}{\m}\\
        \hline
    \end{tabular}
    \caption{First column: OOPLs as estimated from \cite{Brzozowski:LISA-OB} under the assumption of a \SI{10}{\m} fiber length. Second column: a set of OOPLs fulfilling the OB design guideline given by \cref{eq:OpticalBenchDesignGuideline}.}
    \label{tab:OnboardOpticalPathLengths}
\end{table}

\section{Time-delay interferometry with onboard optical delays}\label{sec:TDIwithOOPLs}

\subsection{Brief review of time-delay interferometry}
Time-delay interferometry (TDI) is an on-ground data processing technique for LISA. It time shifts and linearly combines the various interferometric measurements to construct virtual equal-optical-path-length interferometers between the six free-falling test masses. Thus, it mitigates laser frequency noise and OB jitter along the sensitive axis (longitudinal OB jitter). The core\footnote{There are additional processing steps to suppress further noise sources not considered here, notably tilt-to-length couplings \cite{Paczkowski:2022TTL} and clock-related noise \cite{Hartwig:2021Clock}.} TDI algorithm can be divided into three steps \cite{Otto:Thesis}:\par
(1) ISI, TMI, and RFI beatnotes are combined to set up measurements for the interspacecraft-test-mass-to-test-mass separations according to the split interferometry concept. These measurements are free of longitudinal OB jitter. They are called the intermediary TDI $\xi$ variables and are given by
\begin{align}\label{eq:IntermediaryXiWithoutCompensation}
    \xi_{ij} = \text{isi}_{ij}
    + \frac{\text{rfi}_{ij} - \text{tmi}_{ij}}{2}
    + \dot{\textbf{D}}_{ij} \frac{\text{rfi}_{ji} - \text{tmi}_{ji}}{2}.
\end{align}
The ISI measures distance variations between local and distant OBs. The differences between RFI and TMI beatnotes constitute measurements of the OB versus test mass motion on the local and distant SC, respectively.\par
(2) The RFI beatnotes are applied to remove three of six laser noise sources from the intermediary $\xi$ variables. We take the differences between both RFI beatnotes on the same SC to cancel the reciprocal part of the fiber backlink noise. These differences are combined with the intermediary $\xi$ variables (see \cref{eq:IntermediaryXiWithoutCompensation}) to form the intermediary TDI $\eta$ variables, where laser noise contributions of right-handed lasers (those associated with the MOSAs 13, 32, and 21) cancel:
\begin{align}
    \eta_{13} &= \xi_{13}+\frac{\text{rfi}_{12} - \text{rfi}_{13}}{2},\label{eq:IntermediaryEtaRightWithoutCompensation}\\
    \eta_{12} &= \xi_{12} + \dot{\textbf{D}}_{12}\:\frac{\text{rfi}_{21} - \text{rfi}_{23}}{2}.\label{eq:IntermediaryEtaLeftWithoutCompensation}
\end{align}
The remaining $\eta$ variables result from cyclic permutation of the SC indices.\par
(3) The $\eta$ variables are combined to form virtual equal-optical-path-length interferometers, in which laser frequency noise naturally cancels. For example, the second-generation TDI Michelson variable $X_2$ denotes \cite{Bayle:TDI-in-Frequency}
\begin{align}\label{eq:2ndGenerationTdiMichelsonXWithoutCompensation}
	&X_2 =\left(1 - \dot{\textbf{D}}_{121} - \dot{\textbf{D}}_{12131} + \dot{\textbf{D}}_{1312121}\right)\left(\eta_{13} + \dot{\textbf{D}}_{13}\:\eta_{31}\right)\nonumber\\
	&-\left(1 - \dot{\textbf{D}}_{131} - \dot{\textbf{D}}_{13121} + \dot{\textbf{D}}_{1213131}\right)\left(\eta_{12} + \dot{\textbf{D}}_{12}\:\eta_{21}\right),
\end{align}
$Y_2$ and $Z_2$ can be obtained via cyclic permutation of the SC indices.\par
Previous research neglected the coupling of OOPLs in these three steps. Without proper treatment, they cause laser noise residuals. We present analytical models for these residuals in \cref{sec:LaserNoiseResidualsDueToOOPLs}. In this section, we derive a compensation scheme for OOPLs in TDI: We compensate for the corresponding delays by including OOPL delay and advancement operators in the three TDI steps. Below, we refer to the thus updated TDI algorithm as the OOPL compensation scheme (OOPL-CS).

\subsection{Removal of optical bench jitter with OOPLs}\label{sec:RemovalOpticalBenchJitter}
Without compensation, mismatches in the OOPLs between TMI and RFI cause laser noise residuals in the intermediary TDI $\xi$ variables. To compensate for this, we include OOPL operators $\textbf{D}_{\text{a}}$ and $\textbf{D}_{\text{b}}$ in the $\xi$ variables:\footnote{\label{footnote}A priori we do not know whether these operators will turn out to be delay or advancement operators. The delay operator notation is just applied as a placeholder in the derivation.}
\begin{align}\label{eq:IntermediaryXi}
    \xi_{ij} = \text{isi}_{ij}
    + \frac{\textbf{D}_{\text{a}}\text{rfi}_{ij} - \textbf{D}_{\text{b}}\text{tmi}_{ij}}{2}
    + \dot{\textbf{D}}_{ij}\frac{\textbf{D}_{\text{a}}\text{rfi}_{ji} - \textbf{D}_{\text{b}}\text{tmi}_{ji}}{2}.
\end{align}
To derive the required operators $\textbf{D}_{\text{a}}$ and $\textbf{D}_{\text{b}}$ we expand the laser noise terms in the numerators of \cref{eq:IntermediaryXi}:
\begin{align}
    &\:\textbf{D}_{\text{a}}\:\text{rfi}_{ij} - \textbf{D}_{\text{b}}\:\text{tmi}_{ij}\\    =&\left(\textbf{D}_{\text{a}}\textbf{D}_{\text{rfi}}^{\text{adj}} - \textbf{D}_{\text{b}}\textbf{D}_{\text{tmi}}^{\text{adj}}\right)\dot{p}_{ik} - \left(\textbf{D}_{\text{a}}\textbf{D}_{\text{rfi}}^{\text{loc}} - \textbf{D}_{\text{b}}\textbf{D}_{\text{tmi}}^{\text{loc}}\right)\dot{p}_{ij}\nonumber\\
    =&\:\dot{p}_{ik}(\tau - d_{\text{a}} - d_{\text{rfi}}^{\text{adj}}) - \dot{p}_{ik}(\tau - d_{\text{b}} - d_{\text{tmi}}^{\text{adj}})\nonumber\\-&\:\dot{p}_{ij}(\tau - d_{\text{a}} - d_{\text{rfi}}^{\text{loc}}) + \dot{p}_{ij}(\tau - d_{\text{b}}-\: d_{\text{tmi}}^{\text{loc}})\\
    \approx& \left(d_{\text{a}}-d_{\text{b}}+ d_{\text{rfi}}^{\text{loc}}-d_{\text{tmi}}^{\text{loc}} \right) \ddot{p}_{ij}\nonumber\\
    -&\: (d_{\text{a}}-d_{\text{b}}+ d_{\text{rfi}}^{\text{adj}}-d_{\text{tmi}}^{\text{adj}} )\:\ddot{p}_{ik}.\label{eq:DifferenceRfiTmiFirstOrder}
\end{align}
We do not consider the contribution of the next order expansion, which would scale with the OOPL squared and $\dddot{p}$ terms: The two extra time derivatives with respect to $\dot{p}$ give a factor of $(2\pi f)^2$ in terms of amplitude spectral density (ASD); the OOPLs can be approximated with \SI{10}{\nano\s}; consequently, the laser noise contribution of these terms is suppressed by a factor of $10^{-16}\si{\s\tothe{2}}\times (2\pi f)^2$, which is completely negligible across the LISA band.\par
The $\ddot{p}$ terms cannot be dropped so easily. We need to choose the operators $\textbf{D}_{\text{a}}$ and $\textbf{D}_{\text{b}}$ such that the $\ddot{p}$ terms in \cref{eq:DifferenceRfiTmiFirstOrder} cancel. This yields two conditions:
\begin{align}
	d_a - d_b &= d_{\text{tmi}}^{\text{loc}} - d_{\text{rfi}}^{\text{loc}},\\
	d_a - d_b &= d_{\text{tmi}}^{\text{adj}} - d_{\text{rfi}}^{\text{adj}}.
\end{align}
They can be combined to form a guideline for the optical bench design:
\begin{align}
    d_{\text{tmi}}^{\text{adj}} - d_{\text{tmi}}^{\text{loc}} = d_{\text{rfi}}^{\text{adj}} - d_{\text{rfi}}^{\text{loc}},\label{eq:OpticalBenchDesignGuideline}
\end{align}
i.e., the OOPL differences have to match between TMI and RFI. Any deviation from this design guideline will cause a residual laser noise proportional to the deviation, which can be used to formulate a concrete requirement. In fact, the current design values (see the first column in \cref{tab:OnboardOpticalPathLengths}) involve a mismatch of about \SI{2}{\centi\m}. We assess the impact of this mismatch analytically in \cref{sec:LaserNoiseResidualsDueToOOPLs} and numerically in \cref{sec:Results}.\par
If the OB design guideline is fulfilled exactly, we can cancel the $\ddot{p}$ laser noise terms entirely by choosing the delay operators $\textbf{D}_{\text{a}}$ and $\textbf{D}_{\text{b}}$ according to
\begin{align}
    \textbf{D}_{\text{a}}\:\textbf{A}_{\text{b}} &= \textbf{D}_{\text{tmi}}^{\text{loc}}\: \textbf{A}_{\text{rfi}}^{\text{loc}},
\end{align}
which is fulfilled by, e.g., 
\begin{align}
    \textbf{D}_{\text{a}} &= \textbf{D}_{\text{tmi}}^{\text{loc}}\: \textbf{A}_{\text{rfi}}^{\text{loc}},\label{eq:D_a}\\
    \textbf{A}_{\text{b}} &= \mathds{1}.\label{eq:D_b}
\end{align}
$\mathds{1}$ denotes the identity element, which does not change the time argument of the function it is acting on. Thus, we can cancel laser noise terms in the difference between RFI and TMI up to and including the $\ddot{p}$ terms. The intermediary TDI variables $\xi$ can now be written as
\begin{align}
    \xi_{ij} = \text{isi}_{ij} + \dot{N}^{\xi}_{ij},
\end{align}
where $\dot{N}^{\xi}_{ij}$ summarizes backlink, test-mass-acceleration, and readout noise of all constituent beatnotes.\par
The application of the OOPL operators $\textbf{D}_{\text{a}}$ and $\textbf{D}_{\text{b}}$ in \cref{eq:IntermediaryXi} cancels laser frequency noise in the difference between the TMI and RFI beatnotes. It could be argued whether further OOPL operators $\textbf{D}^{\text{ob}}_{\text{c}}$ and $\textbf{D}^{\text{ob}}_{\text{d}}$ are required in \cref{eq:IntermediaryXi} to compensate for the effect of OOPLs in the OB jitter subtraction:
\begin{align}
    \xi_{ij} = \text{isi}_{ij}
    +\: &\textbf{D}^{\text{ob}}_{\text{c}}\:\frac{\textbf{D}_{\text{a}}\text{rfi}_{ij} - \textbf{D}_{\text{b}}\text{tmi}_{ij}}{2}\nonumber\\
    +\: \dot{\textbf{D}}_{ij}\:&\textbf{D}^{\text{ob}}_{\text{d}}\:\frac{\textbf{D}_{\text{a}}\text{rfi}_{ji} - \textbf{D}_{\text{b}}\text{tmi}_{ji}}{2}.
\end{align}
These operators would match the OOPLs between the ISI and the local interferometers. However, OB jitter has a magnitude of \SIrange[]{1}{10}{\nano \m \hertz \tothe{-0.5}} in terms of ASD \cite{Inchauspe:2022SCjitter}, which lies several orders below the laser frequency noise. The coupling of a \SI{3}{\nano\s} OOPL to OB jitter can, thus, be estimated to be in the order of
\begin{align}
    2\pi f \times \SI{3}{\nano\s} \times \SI{10}{\nano \m \hertz \tothe{-0.5}} \approx \SI{e-16}{\m \hertz \tothe{-0.5}} \times \frac{f}{\si{\hertz}},  
\end{align}
which is completely negligible across the LISA band. The factor $2\pi f$ is due the time derivative of the ASD.

\subsection{Reduction to three lasers with OOPLs}\label{sec:ReductionTo3Lasers}
Due to the fiber backlink, local and adjacent OOPLs in the RFI differ by about \SI{10}{\m} (see \cref{tab:OnboardOpticalPathLengths}). If uncompensated, this causes residual laser noise in the intermediary TDI $\eta$ variables. Similarly, OOPL mismatches between ISI and RFI lead to residual laser noise. To compensate for this, we include OOPL operators $\textbf{D}_{\text{c}}$, $\textbf{D}_{\text{d}}$, $\textbf{D}_{\text{e}}$, and $\textbf{D}_{\text{f}}$ in the $\eta$ variables (\cref{footnote} applies):
\begin{align}
    \eta_{13} &= \xi_{13}+\textbf{D}_{\text{e}}\:\frac{\textbf{D}_{\text{c}} \:\text{rfi}_{12} - \textbf{D}_{\text{d}}\:\text{rfi}_{13}}{2},\label{eq:IntermediaryEta1}\\
    \eta_{12} &= \xi_{12}+\dot{\textbf{D}}_{12}\:\textbf{D}_{\text{f}}\:\frac{\textbf{D}_{\text{d}}\:\text{rfi}_{21}-\textbf{D}_{\text{c}}\:\text{rfi}_{23}}{2}.\label{eq:IntermediaryEta2}
\end{align}
We apply $\textbf{D}_{\text{c}}$ to the left-handed RFI beatnotes and $\textbf{D}_{\text{d}}$ to the right-handed ones. $\textbf{D}_{\text{e}}$ and $\textbf{D}_{\text{f}}$ are applied to match the OOPLs between the ISIs and RFIs.\par
To derive the operators $\textbf{D}_{\text{c}}$ and $\textbf{D}_{\text{d}}$ we expand the numerator of \cref{eq:IntermediaryEta1} up to the first order, i.e., up to $\ddot{p}$:
\begin{align}
    &\:\textbf{D}_{\text{c}}\:\text{rfi}_{12} - \textbf{D}_{\text{d}}\:\text{rfi}_{13}\\
    =&\left(\textbf{D}_{\text{c}}\textbf{D}_{\text{rfi}}^{\text{adj}} + \textbf{D}_{\text{d}}\textbf{D}_{\text{rfi}}^{\text{loc}}\right) \dot{p}_{13} - \left(\textbf{D}_{\text{c}}\textbf{D}_{\text{rfi}}^{\text{loc}} + \textbf{D}_{\text{d}}\textbf{D}_{\text{rfi}}^{\text{adj}}\right) \dot{p}_{12}\nonumber\\
    =&\:\dot{p}_{13}(\tau - d_{\text{c}} - d_{\text{rfi}}^{\text{adj}}) + \dot{p}_{13}(\tau - d_{\text{d}} - d_{\text{rfi}}^{\text{loc}})\nonumber\\-&\:\dot{p}_{12}(\tau - d_{\text{c}} - d_{\text{rfi}}^{\text{loc}}) - \dot{p}_{12}(\tau - d_{\text{d}} - d_{\text{rfi}}^{\text{adj}})\\
    \approx &\:2\:\dot{p}_{13} - \ddot{p}_{13}\cdot\left(d_{\text{c}}+d_{\text{d}}+ d_{\text{rfi}}^{\text{loc}}+ d_{\text{rfi}}^{\text{adj}}\right)\nonumber\\
    -&\:2\:\dot{p}_{12} + \ddot{p}_{12}\cdot\left(d_{\text{c}}+d_{\text{d}}+ d_{\text{rfi}}^{\text{loc}}+ d_{\text{rfi}}^{\text{adj}}\right).
\end{align}
The next order terms can be neglected as explained in \cref{sec:RemovalOpticalBenchJitter}. We want to choose the operators $\textbf{D}_{\text{c}}$ and $\textbf{D}_{\text{d}}$ such that the $\ddot{p}$ terms in this expansion cancel. The computation above yields one condition:
\begin{align}
    d_{\text{c}} + d_{\text{d}} + d_{\text{rfi}}^{\text{loc}} + d_{\text{rfi}}^{\text{adj}} = 0,
\end{align}
or in terms of operators
\begin{align}
    \textbf{D}_{\text{c}}\:\textbf{D}_{\text{d}}\:\textbf{D}_{\text{rfi}}^{\text{loc}}\:\textbf{D}_{\text{rfi}}^{\text{adj}} = \mathds{1},\label{eq:ConditionOperatorsTdiStep2}
\end{align}
where the operators commute as the delays are constant. One solution to \cref{eq:ConditionOperatorsTdiStep2} is given by
\begin{align}
	\textbf{D}_{\text{c}} &= \textbf{A}_{\text{rfi}}^{\text{loc}},\label{eq:D_c}\\
	\textbf{D}_{\text{d}} &= \textbf{A}_{\text{rfi}}^{\text{adj}}.\label{eq:D_d}
\end{align}
We then obtain, to first order, our desired result,
\begin{align}
	&\:\textbf{A}_{\text{rfi}}^{\text{loc}} \text{rfi}_{12} - \textbf{A}_{\text{rfi}}^{\text{adj}} \text{rfi}_{13}\\
	=&\:\left(\textbf{A}_{\text{rfi}}^{\text{loc}}\:\textbf{D}_{\text{rfi}}^{\text{adj}} + \textbf{A}_{\text{rfi}}^{\text{adj}}\:\textbf{D}_{\text{rfi}}^{\text{loc}}\right) \dot{p}_{13} -2\:\dot{p}_{12} \\ \approx&\:2\:\dot{p}_{13} - 2\:\dot{p}_{12},
\end{align}
The same result can be derived by considering the numerator of \cref{eq:IntermediaryEta2} instead.\par
Now we derive the operators $\textbf{D}_{\text{e}}$ and $\textbf{D}_{\text{f}}$ to match the OOPLs between ISI and RFI. Choosing $\textbf{D}_{\text{c}}$ and $\textbf{D}_{\text{d}}$ according to \cref{eq:D_c,eq:D_d} allows us to write the right-handed $\eta$ variable as:
\begin{align}
    \eta_{13} =&\: \text{isi}_{13} + \textbf{D}_{\text{e}} \left(\dot{p}_{13} - \:\dot{p}_{12} \right)\\
    =&\:\dot{\textbf{D}}_{13}\:\textbf{D}_{\text{isi}}^{\text{dist}}\:\dot{p}_{31} - \textbf{D}_{\text{e}}\:\dot{p}_{12}\nonumber\\
    -&\:\textbf{D}_{\text{isi}}^{\text{loc}}\:\dot{p}_{13} + \textbf{D}_{\text{e}}\:\dot{p}_{13}.
\end{align}
To cancel the laser frequency noise of the right-handed laser $13$, we set
\begin{align}
    \textbf{D}_{\text{e}}=\textbf{D}_{\text{isi}}^{\text{loc}},\label{eq:D_e}
\end{align}
so that the right-handed $\eta$ variable becomes
\begin{align}
	\eta_{13} = \dot{\textbf{D}}_{13}\:\textbf{D}_{\text{isi}}^{\text{dist}}\:\dot{p}_{31} - \textbf{D}_{\text{isi}}^{\text{loc}}\:\dot{p}_{12}.
\end{align}
With \cref{eq:D_c,eq:D_d} the left-handed intermediary $\eta$ variable can be written as
\begin{align}
    \eta_{12} =&\: \text{isi}_{12} + \dot{\textbf{D}}_{12}\:\textbf{D}_{\text{f}}\:\left(\dot{p}_{23} - \dot{p}_{21}\right)\\
    =&\:\dot{\textbf{D}}_{12}\:\textbf{D}_{\text{f}} \: \dot{p}_{23} - \textbf{D}_{\text{isi}}^{\text{loc}}\:\dot{p}_{12}\nonumber\\
    +&\:\dot{\textbf{D}}_{12}\:\textbf{D}_{\text{isi}}^{\text{dist}}\: \dot{p}_{21} - \dot{\textbf{D}}_{12}\:\textbf{D}_{\text{f}} \: \dot{p}_{21}.
\end{align}
We want to cancel the right-handed laser 21, so we set
\begin{align}
    \textbf{D}_{\text{f}}=\textbf{D}_{\text{isi}}^{\text{dist}},\label{eq:D_f}
\end{align}
and the left-handed $\eta$ variable becomes
\begin{align}
    \eta_{12} &=\dot{\textbf{D}}_{12}\:\textbf{D}_{\text{isi}}^{\text{dist}}\:\dot{p}_{23} - \textbf{D}_{\text{isi}}^{\text{loc}}\:\dot{p}_{12}.
\end{align}
Thus, we cancel the laser frequency noise contributions of right-handed lasers (up to and including second order $\ddot{p}$) by including the OOPL operators $\textbf{D}_{\text{c}}$, $\textbf{D}_{\text{d}}$, $\textbf{D}_{\text{e}}$, and $\textbf{D}_{\text{f}}$ as defined in \cref{eq:D_c,eq:D_d,eq:D_e,eq:D_f} in the $\eta$ variables \cref{eq:IntermediaryEta1,eq:IntermediaryEta2}.

\subsection{Laser-noise-free TDI combinations with OOPLs}
Laser-noise-free TDI combinations like the second-generation TDI Michelson variable $X_2$ (see \cref{eq:2ndGenerationTdiMichelsonXWithoutCompensation}) are linear combinations of the $\eta$ variables delayed by interspacecraft delay operators $\dot{\textbf{D}}_{ij}$. While $\dot{\textbf{D}}_{ij}$ just accounts for the interspacecraft signal propagation time (defined between the PBSs on receiving and emitting SC), the $\eta$ variables are defined at the combining BS of the ISI (we neglect subsequent common delays due to analog and digital signal processing). Hence, the OOPLs in the ISI cause residual laser frequency noise if uncompensated.\par
\cite{Reinhardt:RangingSensorFusion} considers a TDI toy model to derive an operator $\dot{\mathcal{D}}_{ij}$ that accounts for onboard delays. 
This TDI delay operator then replaces $\dot{\textbf{D}}_{ij}$ in the TDI combinations. However, the TDI steps 1 and 2 are neglected there, and the TDI delay operator is directly computed from the ISI beatnotes. We, therefore, revisit that toy model here and rebuild it upon the intermediary TDI $\eta$ variables as derived in \cref{sec:ReductionTo3Lasers}.\par
We combine $\eta_{12}$ and $\eta_{21}$ and focus on canceling the noise of laser $12$:
\begin{align}
    \dot{\mathcal{D}}_{21} \: \eta_{12} + 	\eta_{21}=&\:\dot{\mathcal{D}}_{21}\left( \dot{\textbf{D}}_{12}\:\textbf{D}_{\text{isi}}^{\text{dist}}\:\dot{p}_{23} - \textbf{D}_{\text{isi}}^{\text{loc}}\:\dot{p}_{12}\right)\nonumber\\
    +&\:\dot{\textbf{D}}_{21}\:\textbf{D}_{\text{isi}}^{\text{dist}}\:\dot{p}_{12} - \textbf{D}_{\text{isi}}^{\text{loc}}\:\dot{p}_{23} ,\\
    =\: \Big( \dot{\textbf{D}}_{21}\:\textbf{D}_{\text{isi}}^{\text{dist}} - \dot{\mathcal{D}}_{21}&\:\textbf{D}_{\text{isi}}^{\text{loc}} \Big)\dot{p}_{12} + (...)\:\dot{p}_{23}.
\end{align}
To cancel $\dot{p}_{12}$, we need to choose $\dot{\mathcal{D}}_{21}$ such that the first bracket vanishes, i.e.,
\begin{align}
    \dot{\textbf{D}}_{21}\:\textbf{D}_{\text{isi}}^{\text{dist}} = \dot{\mathcal{D}}_{21}&\:\textbf{D}_{\text{isi}}^{\text{loc}}.
\end{align}
Multiplying this expression with $\textbf{A}_{\text{isi}}^{\text{loc}}$ from the right yields the TDI delay operator
\begin{align}\label{eq:TDI_Delay}
    \dot{\mathcal{D}}_{21} = \dot{\textbf{D}}_{21}\:\textbf{D}_{\text{isi}}^{\text{dist}}\:\textbf{A}_{\text{isi}}^{\text{loc}},
\end{align}
which accounts for OOPLs in the ISI. It replaces $\dot{\textbf{D}}$ in the laser-noise-free TDI combinations, e.g., \cref{eq:2ndGenerationTdiMichelsonXWithoutCompensation} becomes
\begin{align}
	&X_2 =\left(1 - \dot{\mathcal{D}}_{121} - \dot{\mathcal{D}}_{12131} + \dot{\mathcal{D}}_{1312121}\right)\left(\eta_{13} + \dot{\mathcal{D}}_{13}\:\eta_{31}\right)\nonumber\\
	&-\left(1 - \dot{\mathcal{D}}_{131} - \dot{\mathcal{D}}_{13121} + \dot{\mathcal{D}}_{1213131}\right)\left(\eta_{12} + \dot{\mathcal{D}}_{12}\:\eta_{21}\right).
\end{align}
Other TDI combinations can be updated analogously by replacing the $\dot{\textbf{D}}$ operators with $\dot{\mathcal{D}}$ operators.

\section{Impact of onboard optical Delays}\label{sec:LaserNoiseResidualsDueToOOPLs}
In the previous section, we derived an OOPL compensation scheme for TDI, which includes OOPL delay and advancement operators in the TDI equations. In this section, we derive analytical models for the coupling of OOPLs in TDI if uncompensated. We focus on the first two steps (computation of the intermediary TDI variables $\xi$ and $\eta$). In the third TDI step, OOPLs act as biases; their effect has been studied in \cite{Staab:LaserNoiseResiduals,Staab:Thesis}.\par
We start with the intermediary TDI $\xi$ variables (see \cref{eq:IntermediaryXiWithoutCompensation}). Now we do not compensate OOPLs with the operators $\textbf{D}_{\text{a}}$ and $\textbf{D}_{\text{b}}$ as in \cref{eq:IntermediaryXi}. Expanding the difference between RFI and TMI beatnotes to first order gives:
\begin{align}
    \text{rfi}_{ij} - \text{tmi}_{ij}
    =&\:\dot{p}_{ik}(\tau -d_{\text{rfi}}^{\text{adj}}) - \dot{p}_{ij}(\tau - d_{\text{rfi}}^{\text{loc}})\nonumber\\ -&\left(\dot{p}_{ik}(\tau -d_{\text{tmi}}^{\text{adj}}) - \dot{p}_{ij}(\tau -d_{\text{tmi}}^{\text{loc}})\right)\\
    \approx&\:\dot{p}_{ik} - d_{\text{rfi}}^{\text{adj}} \: \ddot{p}_{ik} - \dot{p}_{ij} + d_{\text{rfi}}^{\text{loc}} \: \ddot{p}_{ij}\nonumber\\-&\:\dot{p}_{ik} + d_{\text{tmi}}^{\text{adj}} \: \ddot{p}_{ik} + \dot{p}_{ij} - d_{\text{tmi}}^{\text{loc}} \: \ddot{p}_{ij}\\
    =&\:d_{\Delta}^{\text{loc}} \: \ddot{p}_{ij} - d_{\Delta}^{\text{adj}} \: \ddot{p}_{ik},\label{eq:DifferenceRfiTmiWithoutCompensation}
\end{align}
where we introduce the abbreviations
\begin{align}
    d_{\Delta}^{\text{loc}}&=d_{\text{rfi}}^{\text{loc}} - d_{\text{tmi}}^{\text{loc}},\\
    d_{\Delta}^{\text{adj}}&=d_{\text{rfi}}^{\text{adj}} - d_{\text{tmi}}^{\text{adj}}
\end{align}
for mismatches between OOPLs in TMI and RFI. Inserting \cref{eq:DifferenceRfiTmiWithoutCompensation} into the $\xi$ variables yields:
\begin{align}
    \xi_{ij} =&\:\text{isi}_{ij} + \frac{d_{\Delta}^{\text{loc}} \ddot{p}_{ij} - d_{\Delta}^{\text{adj}} \ddot{p}_{ik}}{2} + \dot{\textbf{D}}_{ij}\frac{d_{\Delta}^{\text{loc}} \ddot{p}_{ji} - d_{\Delta}^{\text{adj}} \ddot{p}_{jk}}{2}\\
    =&\:\text{isi}_{ij} + C^{\xi}_{ij},\\
    C^{\xi}_{ij} \vcentcolon=&\:\frac{d_{\Delta}^{\text{loc}} \: \ddot{p}_{ij} - d_{\Delta}^{\text{adj}} \: \ddot{p}_{ik}}{2} + \dot{\textbf{D}}_{ij}\frac{d_{\Delta}^{\text{loc}} \: \ddot{p}_{ji} - d_{\Delta}^{\text{adj}} \: \ddot{p}_{jk}}{2},
\end{align}
where $C^{\xi}_{ij}$ is the correction term due to OOPLs.\par
We proceed with the intermediary TDI $\eta$ variables (see \cref{eq:IntermediaryEtaRightWithoutCompensation,eq:IntermediaryEtaLeftWithoutCompensation}). Now we do not compensate OOPLs with the operators $\textbf{D}_{\text{c}}$ to $\textbf{D}_{\text{f}}$ as in \cref{eq:IntermediaryEta1,eq:IntermediaryEta2}. We expand the ISI and RFI beatnotes to first order:
\begin{align}
    \text{rfi}_{ij} &\approx \dot{p}_{ik} - d_{\text{rfi}}^{\text{adj}} \: \ddot{p}_{ik} - \left(\dot{p}_{ij} - d_{\text{rfi}}^{\text{loc}} \: \ddot{p}_{ij}\right),\label{eq:RfiToFirstOrder}\\
    \text{isi}_{ij} &\approx \dot{\textbf{D}}_{ij}\left(\dot{p}_{ji} - d_{\text{isi}}^{\text{dist}} \: \ddot{p}_{ji}\right) - \left(\dot{p}_{ij} - d_{\text{isi}}^{\text{loc}} \: \ddot{p}_{ij}\right).\label{eq:IsiToFirstOrder}
\end{align}
Inserting \cref{eq:RfiToFirstOrder,eq:IsiToFirstOrder} into \cref{eq:IntermediaryEtaLeftWithoutCompensation} for the right-handed $\eta$ variables yields:
\begin{align}
    \eta_{12} =&\: \text{isi}_{12} + \dot{\textbf{D}}_{12}\frac{\text{rfi}_{21} - \text{rfi}_{23}}{2} + C^{\xi}_{12}\\
    =&\: \bar{\eta}_{12} + C^{\xi}_{12} + C^{\eta}_{12},\label{eq:TDIetaWithCorrection1}\\
    \bar{\eta}_{12}\vcentcolon=&\: \dot{\textbf{D}}_{12}\:\dot{p}_{23} - \dot{p}_{12},\\
    C^{\eta}_{12}\vcentcolon=&\:\frac{d_{\text{rfi}}^{\text{adj}} + d_{\text{rfi}}^{\text{loc}}}{2}\:\dot{\textbf{D}}_{12}\:\left(\ddot{p}_{21} - \ddot{p}_{23}\right)\nonumber\\
    +&\:d_{\text{isi}}^{\text{loc}} \: \ddot{p}_{12} - d_{\text{isi}}^{\text{dist}} \: \dot{\textbf{D}}_{12}\:\ddot{p}_{21}\\
    \approx&\: \frac{d_{\text{rfi}}^{\text{adj}}}{2}\dot{\textbf{D}}_{12}\left(\ddot{p}_{21} - \ddot{p}_{23}\right),\label{eq:C-Eta-Left-Approximated}
\end{align}
where $\bar{\eta}$ denotes the common $\eta$ variable without OOPLs. $C^{\eta}$ is the correction term due to OOPLs in TDI step 2. In \cref{eq:C-Eta-Left-Approximated}, we neglect the contribution of local and distant OOPLs and focus on the adjacent ones, which dominate due to the \SI{10}{\m} backlink fiber. For the right-handed intermediary $\eta$ variables, we analogously obtain
\begin{align}
    \eta_{13} =&\: \text{isi}_{13} + \frac{\text{rfi}_{12} - \text{rfi}_{13}}{2} + C^{\xi}_{13}\\
    =&\: \bar{\eta}_{13} + C^{\xi}_{13} + C^{\eta}_{13},\label{eq:TDIetaWithCorrection2}\\
    \bar{\eta}_{13}\vcentcolon=&\:\dot{\textbf{D}}_{13}\:\dot{p}_{31} - \dot{p}_{12},\\
    C^{\eta}_{13}\vcentcolon=&\: \frac{d_{\text{rfi}}^{\text{adj}} + d_{\text{rfi}}^{\text{loc}}}{2}\left(\ddot{p}_{12} - \ddot{p}_{13}\right)\nonumber\\
    +&\:d_{\text{isi}}^{\text{loc}} \: \ddot{p}_{13} - d_{\text{isi}}^{\text{dist}} \: \dot{\textbf{D}}_{13}\:\ddot{p}_{31}\\
    \approx&\: \frac{d_{\text{rfi}}^{\text{adj}}}{2}\left(\ddot{p}_{12} - \ddot{p}_{13}\right).\label{eq:C-Eta-Right-Approximated}
\end{align}
Again, we drop local and distant OOPLs and focus on the adjacent ones, which are 1 order of magnitude higher.\par
We plug the above computed $\eta$ variables (see \cref{eq:TDIetaWithCorrection1,eq:TDIetaWithCorrection2}) into the second-generation TDI Michelson variable $X_2$ (see \cref{eq:2ndGenerationTdiMichelsonXWithoutCompensation}) and compute the laser noise residuals associated with the correction terms $C^{\xi}$ and $C^{\eta}$. We consider the equal arm approximation so that we can drop the indices of interspacecraft delay operators. We further assume constant arms taking $\dot{\textbf{D}} = \textbf{D}$. For chained delay operators in the equal arm approximation, we introduce the shorthand notation
\begin{align}
    \underbrace{\:\textbf{D}\:\textbf{D}\: ... \:\textbf{D}\:}_{n} =\vcentcolon \textbf{D}^{n}.
\end{align}
The correction terms $C^{\xi}$ and $C^{\eta}$ are additive. Consequently, we can write $X_2$ as
\begin{align}
    X_2 = \bar{X}_2 + X^{\xi} + X^{\eta}.
\end{align}
$\bar{X}_2$ is the common laser noise canceling TDI variable.\par
$X^{\xi}$ is the laser noise residual associated with the correction term $C^{\xi}$, which can be computed to be
\begin{widetext}
    \begin{align}
        X^{\xi} &= (1 - \textbf{D}^2 - \textbf{D}^4 + \textbf{D}^6)\left(C^{\xi}_{13} + \textbf{D}\: C^{\xi}_{31} - C^{\xi}_{12} - \textbf{D}\: C^{\xi}_{21}\right)\\
        &= (1 - \textbf{D}^2 - \textbf{D}^4 + \textbf{D}^6)
        \left( \frac{d_{\Delta}^{\text{loc}} + d_{\Delta}^{\text{adj}}}{2} \left(1 + \textbf{D}^2\right) \left(\ddot{p}_{13} - \ddot{p}_{12}\right)
        + d_{\Delta}^{\text{loc}} \: \textbf{D} \left(\ddot{p}_{31} - \ddot{p}_{21}\right) + d_{\Delta}^{\text{adj}} \: \textbf{D} \left(\ddot{p}_{23} - \ddot{p}_{32}\right) \right).\label{eq:TdiXxiCompleteResidual}
    \end{align}
\end{widetext}
It scales with $d_{\Delta}^{\text{loc}}$ and $d_{\Delta}^{\text{adj}}$, which amount to \SI{5}{\centi\m} and \SI{7}{\centi\m} according to the current OB design (see \cref{tab:OnboardOpticalPathLengths}). The first term in \cref{eq:TdiXxiCompleteResidual} vanishes for laser-locking configurations that lock the lasers 12 and 13 directly onto each other, assuming sufficient gain of the locking control loop. We discuss the effect of different laser locking configurations in \cref{app:LaserLocking}.\par
If the OB fulfills the OB design guideline (see \cref{eq:OpticalBenchDesignGuideline}), $X^{\xi}$ can be canceled completely by including OOPL delay operators into the intermediary $\xi$ variables (see \cref{sec:RemovalOpticalBenchJitter}). In the case of mismatches between $d_{\Delta}^{\text{loc}}$ and $d_{\Delta}^{\text{adj}}$, $X^{\xi}$ can not be canceled completely. The laser noise residual due to such mismatches can be computed from \cref{eq:TdiXxiCompleteResidual}: We choose the operators $\textbf{D}_{\text{a}}$ and $\textbf{D}_{\text{b}}$ according to \cref{eq:D_a,eq:D_b}, i.e., we match them with $d_{\Delta}^{\text{loc}}$. This cancels the terms in \cref{eq:TdiXxiCompleteResidual} that are proportional to $d_{\Delta}^{\text{loc}}$. However, for the $d_{\Delta}^{\text{adj}}$ terms, we then obtain laser noise residuals that scale with the OB mismatch
\begin{align}
    \Delta_{\text{ob}} = d_{\text{tmi}}^{\text{adj}} - d_{\text{tmi}}^{\text{loc}} - \left(d_{\text{rfi}}^{\text{adj}} - d_{\text{rfi}}^{\text{loc}}\right),
\end{align}
which amounts to \SI{2}{\centi\m} according to the current OB design. The associated laser noise residual is given by
\begin{align}
    X^{\xi}_{\Delta_{\text{ob}}} =&\: \frac{\Delta_{\text{ob}}}{2}\:(1 - 2\textbf{D}^4 + \textbf{D}^8)\left(\ddot{p}_{13} - \ddot{p}_{12}\right)\\
    +&\:\Delta_{\text{ob}}\:(\textbf{D} - \textbf{D}^3 - \textbf{D}^5 + \textbf{D}^7)\left(\ddot{p}_{23} - \ddot{p}_{32}\right),\label{eq:XXi}
\end{align}
it scales with the OB mismatch $\Delta_{\text{ob}}$.\par
$X^{\eta}$ is the laser noise residual associated with the correction term $C^{\eta}$, which can be computed to be
\begin{align}
    X^{\eta} &= (1 - \textbf{D}^2 - \textbf{D}^4 + \textbf{D}^6)\nonumber\\
    &\hspace{1cm}\left(C^{\eta}_{13} + \textbf{D}\: C^{\eta}_{31} - C^{\eta}_{12} - \textbf{D}\: C^{\eta}_{21}\right)\\
    &= \frac{d_{\text{rfi}}^{\text{adj}}}{2} \left(1 - 2\textbf{D}^2 + 2\textbf{D}^6 - \textbf{D}^8\right) \left(\ddot{p}_{12} - \ddot{p}_{13}\right).\label{eq:XEta}
\end{align}
The laser noise residual $X^{\eta}$ scales with $d_{\text{rfi}}^{\text{adj}} \approx \SI{10}{\m}$. Note that laser noise terms in the above equations will be correlated due to the laser locking control loops governing the frequency relationships. For example, assuming sufficient gain in the control loop, the residual in \cref{eq:XEta} will cancel exactly if the two local lasers are locked together. We discuss the effect of different laser locking configurations in \cref{app:LaserLocking}.\par
We now derive the ASDs associated with the laser noise residuals $X^{\xi}_{\Delta_{\text{ob}}}$ and $X^{\eta}$. The ASD of a variable $X$ is the square root of the power spectral density (PSD), labeled $S_X$. The PSD, or more generally the cross-spectral density (CSD) $S_{XY}$ if two different variables $X,\:Y$ are considered, can be computed according to \cite{Risken:PSD}
\begin{align}
    S_{XY}(\omega) =& \int \text{d} \omega^{\prime}\:\langle\tilde{X}(\omega) \: \tilde{Y}^{\star}(\omega^{\prime})\rangle,
    \\
    S_{X}(\omega) \vcentcolon=& S_{XX}(\omega).
    \label{eq:PSD}
\end{align}
The tilde denotes the Fourier transform, and the bracket indicates that we must take the expectation value. With the time delay property of the Fourier transform
\begin{align}
    \textbf{D}\tilde{X}(\omega) = e^{-i \omega d} \tilde{X}(\omega),
\end{align}
we can compute $\tilde{X}^{\xi}_{\Delta_{\text{ob}}}$ and $\tilde{X}^{\eta}$ to be:
\begin{widetext}
    \begin{align}
        \tilde{X}^{\xi}_{\Delta_{\text{ob}}}(\omega) &= \frac{\Delta_{\text{ob}}}{2}(1 - 2e^{-4i \omega d} + e^{-8i \omega d}) (\ddot{\tilde{p}}_{13} - \ddot{\tilde{p}}_{12})(\omega)
        +\:\Delta_{\text{ob}}(e^{-i \omega d} - e^{-3i \omega d} - e^{-5i \omega d} + e^{-7i \omega d}) (\ddot{\tilde{p}}_{23} - \ddot{\tilde{p}}_{32})(\omega),\\
        \tilde{X}^{\eta}(\omega) &= \frac{d_{\text{rfi}}^{\text{adj}}}{2}(1 - 2e^{-2i \omega d} + 2e^{-6i \omega d} - e^{-8i \omega d}) (\ddot{\tilde{p}}_{12} - \ddot{\tilde{p}}_{13})(\omega).
    \end{align}
\end{widetext}
Plugging $\tilde{X}^{\eta}$ into \cref{eq:PSD} yields:
\begin{align}
    S_{X^{\eta}}(\omega) &= (4 \: \omega \: d_{\text{rfi}}^{\text{adj}})^2 \sin \left( \omega d\right)^4 \sin\left(2\omega d\right)^2\nonumber\\
    &\hspace{0.7cm}\Big{(}S_{\dot{p}_{12}}(\omega) + S_{\dot{p}_{13}}(\omega) - 2 S_{\dot{p}_{12}\:\dot{p}_{13}}(\omega)\Big{)},
\end{align}
where $S_{\dot{p}_{12}\:\dot{p}_{13}}$ denotes the laser frequency noise cross-spectral density of the lasers 12 and 13. To simplify, we now assume that all lasers are uncorrelated, i.e., $S_{\dot{p}_{12}\:\dot{p}_{13}}(\omega)=0$, and that they have the same ASD, which we denote by $\sqrt{S_{\dot{p}}}$. This leads to the following expression for the ASD of $X^{\eta}$:
\begin{align}\label{eq:ASDXeta}
    \sqrt{\text{S}_{X^{\eta}}(\omega)}
    =\sqrt{32}\:\omega\:d_{\text{rfi}}^{\text{adj}}\:\sqrt{S_{\dot{p}}(\omega)} \sin \left( \frac{\omega L}{c}\right)^2 \sin\left(\frac{2\omega L}{c} \right),
\end{align}
where we expressed the delay in terms of the LISA arm length $L\approx \SI{2.5}{\giga\m}$. This ASD depends on the laser noise ASD $\sqrt{S_{\dot{p}}(\omega)}$ and on the adjacent OOPL in the RFI $d_{\text{rfi}}^{\text{adj}}$. Analogously, we plug $\tilde{X}^{\xi}_{\Delta_{\text{ob}}}$ into \cref{eq:PSD}, which yields an expression for the ASD of $X^{\xi}_{\Delta_{\text{ob}}}$:
\begin{align}\label{eq:ASDImpact2cmMismatch}
    &\sqrt{\text{S}_{X^{\xi}}(\omega)}
    =\:\omega \:\Delta_{\text{ob}}\:\sqrt{S_{\dot{p}}(\omega)} \nonumber\\
    &\hspace{0.4cm}\sqrt{8 \sin \left(\frac{2\omega L}{c}\right)^4 + 32 \sin \left(\frac{\omega L}{c}\right)^2\:\sin\left(\frac{2\omega L}{c}\right)^2}.
\end{align}
Again, we assumed uncorrelated lasers, i.e., pairwise vanishing laser noise cross-spectral densities. This ASD depends on the laser noise ASD $\sqrt{S_{\dot{p}}(\omega)}$ and on the OB mismatch $\Delta_{\text{ob}}$.

\section{Results}\label{sec:Results}
\begin{figure*}
	\begin{center}
		\includegraphics[width=1.0\textwidth]{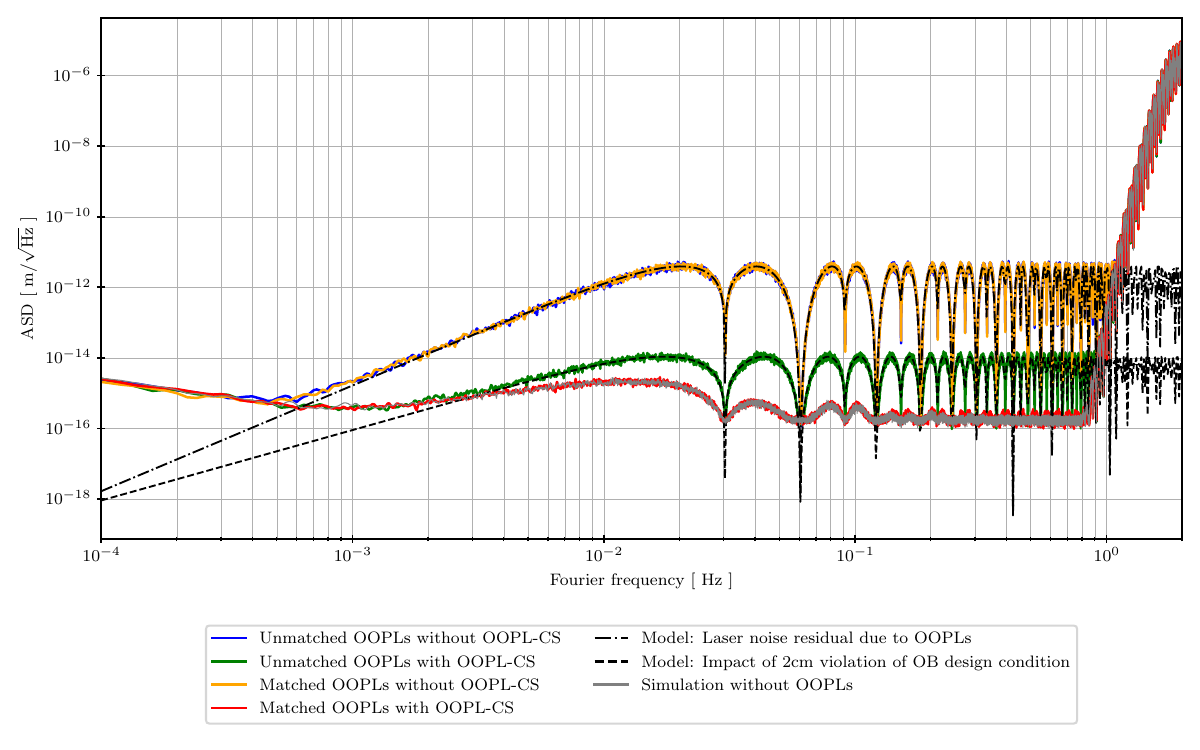}
	\end{center}
	\caption{Residual laser noise in $X_2$ for three simulations with different OOPLs. Grey: Simulation without OOPLs. Blue and green: Simulation with the current design values. Orange and red: Simulation with matched OOPLs. Blue and orange: Without application of the OOPL-CS. Red and green: With the application of the OOPL-CS. Black dash-dotted: Analytical model for the laser noise residual due to OOPLs in \si{\m \hertz \tothe{-0.5}} (compare with \cref{eq:ASDXeta}). Black dashed: Analytical model for the impact of the \SI{2}{\centi\m} violation of the OB design guideline in \si{\m \hertz \tothe{-0.5}} (compare with \cref{eq:ASDImpact2cmMismatch}).}\label{fig:ResultsJustLaserNoise}
\end{figure*}
\begin{figure*}
	\begin{center}
		\includegraphics[width=1.0\textwidth]{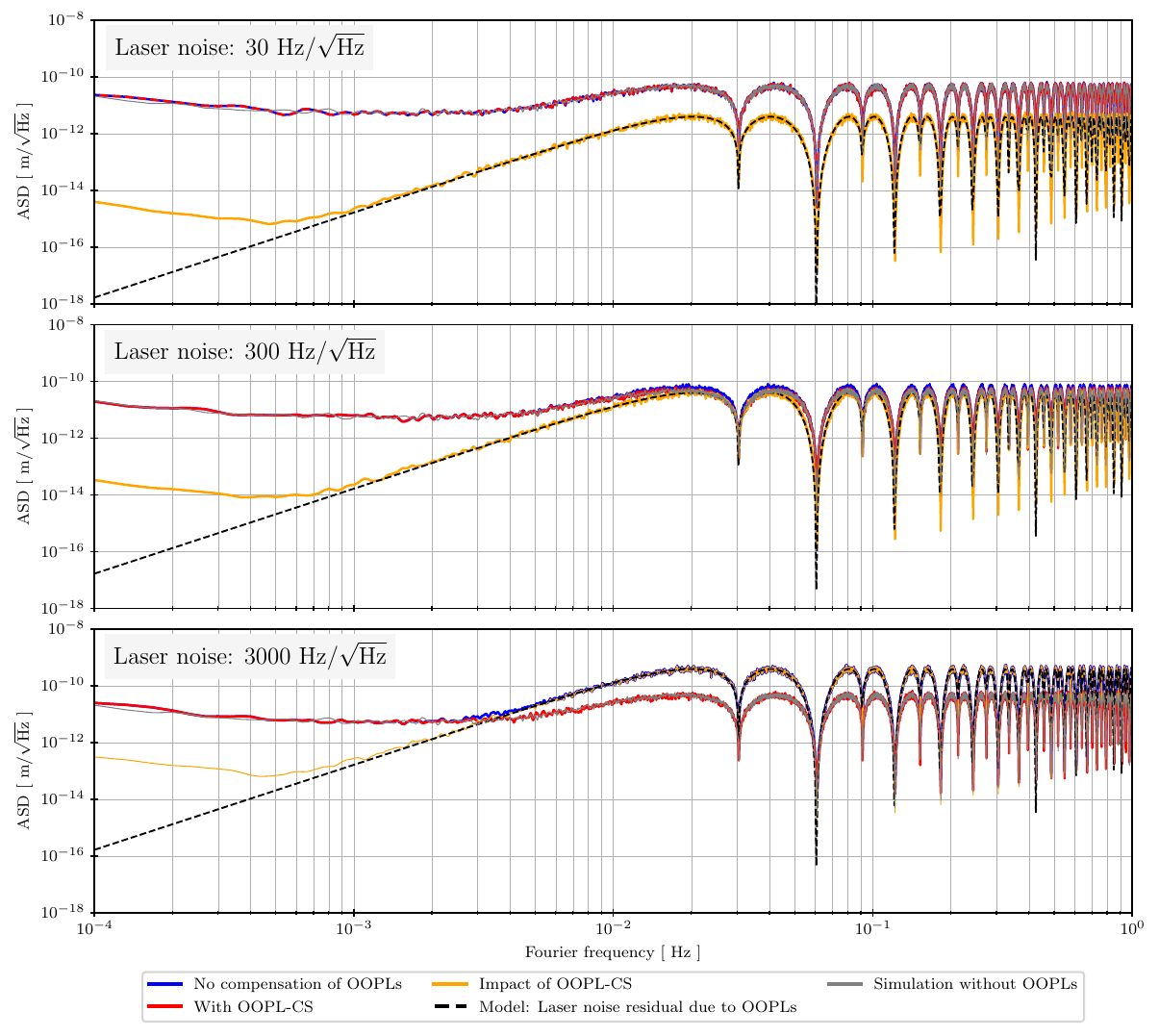}
	\end{center}
	\caption{Simulation including secondary noises. We consider three different laser noise levels. Blue: Residual noise in $X_2$ without OOPL-CS. Red: Residual noise in $X_2$ with OOPL-CS. Orange: Impact of the compensation. Grey: Residual noise in $X_2$ for a simulation without OOPLs. Black dashed: Analytical model for the OOPL-related laser noise residual.}\label{fig:ResultsWithSecondaryNoises}
\end{figure*}
In this section, we verify and demonstrate the performance improvement due to the OOPL-CS (see \cref{sec:TDIwithOOPLs}) via numerical simulations using PyTDI \cite{Staab:PyTDI}. We numerically assess the laser noise residuals associated with OOPLs and the impact of the \SI{2}{\centi\m} violation of the OB design guideline (see \cref{eq:OpticalBenchDesignGuideline}). Finally, we compare the results with the analytical models derived in \cref{sec:LaserNoiseResidualsDueToOOPLs}.\par
We conduct two studies: In the first one (\cref{sec:ResultsJustLaserNoise}), we just consider white laser frequency noise with the ASD
\begin{align}
    \sqrt{S_{\dot{p}}(f)} = \SI{30}{\hertz \hertz\tothe{-0.5}}.\label{eq:ASDLaserFrequencyNoise}
\end{align}
In the second one (\cref{sec:ResultsAlsoSecondaryNoises}), we add secondary noises and investigate the impact of underperforming lasers with laser noise ASDs of \num{300} and \SI{3000}{\hertz \hertz\tothe{-0.5}}. In both studies, we consider telemetry data simulated with LISA Instrument \cite{Bayle:LisaInstrument} using an orbit file provided by ESA~\cite{Martens:2021TrajectoryDesign}, interfaced via LISA Orbits \cite{Bayle:LisaOrbits}. We do not apply laser locking. Its impact is discussed in \cref{app:LaserLocking}. We neglect clock noise, clock desynchronizations, and any effects related to tilt-to-length couplings. Furthermore, we assume the intersatellite ranges to be perfectly known and neglect ranging noise. A framework to obtain accurate and precise estimates for the intersatellite ranges is provided in \cite{Reinhardt:RangingSensorFusion}.

\subsection{Simulation with just laser frequency noise}\label{sec:ResultsJustLaserNoise}
We perform two simulations: In the first one, we apply the current OOPL design values (first column in \cref{tab:OnboardOpticalPathLengths}). In the second simulation, we consider matched OOPLs (second column in \cref{tab:OnboardOpticalPathLengths}), i.e., we add \SI{2}{\centi\m} to $d_{\text{rfi}}^{\text{adj}}$ so that \cref{eq:OpticalBenchDesignGuideline} is fulfilled exactly. In both simulations, we add manufacturing asymmetries in the order of \SI{100}{\micro\m} as described in \cref{app:ManufacturingAsymmetries}. We compute the second-generation TDI X Michelson variables with and without the OOPL-CS for both simulations. For comparison, we consider a third simulation without any OOPLs.\par
Fig. \ref{fig:ResultsJustLaserNoise} shows the corresponding ASDs as displacement noise in \si{\m \hertz \tothe{-0.5}}. The steep slopes above \SI{1}{\hertz} are caused by aliasing and interpolation errors \cite{Staab:LaserNoiseResiduals}. In the case of matched OOPLs, the results with and without the OOPL-CS are plotted in red and orange, respectively. The grey line shows the results of the simulation without OOPLs. For matched OOPLs, the OOPL-CS suppresses laser frequency noise by about 4 orders of magnitude (red versus orange). It reaches the performance of the simulation without OOPLs (red versus grey). The black dash-dotted line shows the analytical model for the
impact of OOPLs (see \cref{eq:ASDXeta}), which focuses on the dominating adjacent OOPLs. Above \SI{1}{\milli\hertz}, the model agrees with the plots for simulations without the OOPL-CS (black dash-dotted versus blue and orange). Hence, in this simulation with just laser frequency noise, the performance above \SI{1}{\milli\hertz} is limited by the adjacent OOPLs. Note that the increase in the low frequencies is slightly misleading: While the plot is displayed in units of \si{\meter \hertz\tothe{-0.5}}, the actual simulation was done in units of frequency, i.e., \si{\hertz \hertz\tothe{-0.5}}. In this raw data, this deviation corresponds to a white noise floor consistent with previous simulation results from the same software \cite{Bayle:TDI-in-Frequency}.\par
In the case of the current OOPL design values, the results with and without the OOPL-CS are plotted in green and blue, respectively. For these unmatched OOPLs, the OOPL-CS reduces the laser frequency noise by more than 2 orders of magnitude (green versus blue). Here, its performance is limited by the laser noise residual associated with the \SI{2}{\centi\m} violation of the OB design guideline. The analytical model of this \SI{2}{\centi\m} mismatch (see \cref{eq:ASDImpact2cmMismatch}) is represented by the black dashed line and agrees with the numerical result (black dashed versus green).

\subsection{Simulation including secondary noises}\label{sec:ResultsAlsoSecondaryNoises}
We consider a more realistic scenario and add readout, test-mass acceleration, and backlink noise as specified in \cite{Bayle:LISA-Simulation}. \Cref{fig:ResultsWithSecondaryNoises} shows the performance of the OOPL-CS for different laser noise levels in this environment: In the upper plot we consider realistic laser noise (see \cref{eq:ASDLaserFrequencyNoise}); the central and lower plots represent the case of underperforming lasers with laser noise ASDs of \num{300} and \SI{3000}{\hertz \hertz\tothe{-0.5}}. In all plots, the red and blue curves show the noise in $X_2$ with and without the OOPL-CS, and the grey plot is a simulation without OOPLs. In orange, we depict the impact of the OOPL-CS, i.e., the ASD of the difference between $X_2$ with and without application of the OOPL-CS. The black dashed line represents the analytical model of OOPLs (see \cref{eq:ASDXeta}).\par
For all laser frequency noise levels, the impact of the OOPL-CS matches the analytical model above \SI{1}{\milli\hertz} (orange versus black dashed). In the realistic case (upper plot), the impact of OOPLs reaches a few \si{\pico \m \hertz\tothe{-0.5}} but remains below the secondary noise levels (orange and black dashed versus grey). Hence, OOPLs do not limit the performance here and the OOPL-CS does not yield an improvement (red versus blue). Nonetheless, every noise above \SI{1}{\pico \m \hertz\tothe{-0.5}} should be subject to a thorough evaluation, and it is advisable to consider the application of the OOPL-CS.\par
In the scenario of underperforming lasers, the subtraction of OOPL-related laser noise residuals becomes critical: They compete with the secondary noise levels for increased laser noise of \SI{300}{\hertz \hertz\tothe{-0.5}} (central plot); in the case of \SI{3000}{\hertz \hertz\tothe{-0.5}} (lower plot), the impact of OOPLs surpasses the secondary noise levels by about 2 orders of magnitude (orange and black dashed versus grey). In both cases, the OOPL-CS successfully removes the OOPL-related laser noise residuals (blue versus red and red versus grey), thus leading to an improvement in performance.


\section{Conclusion}\label{sec:Conclusion}
TDI applies estimates for the interspacecraft distances to construct equal-optical-path-length interferometers from the LISA interferometric measurements. In a realistic LISA simulation, which does not simplify the LISA satellites as point masses, onboard delays become a non-negligible part of these equal-optical-path-length interferometers. Previous research studied TDI in the context of onboard delays after the combining BSs, which are common to both interfering beams (electronic delays in the analog backend, etc.) \cite{Euringer:FrontendAndModulationDelays}. Onboard delays before the combining BSs due to onboard optical path lengths (OOPLs) are non-common between both interfering beams and require a completely different treatment. \cite{Reinhardt:RangingSensorFusion} briefly discussed OOPLs in the ISI but neglected the OOPLs in TMI and RFI, which dominate due to the relatively long fiber backlink. We here extend this work to OOPLs in all interferometers, and substantiate it with analytical models and numerical simulations.\par
We derive an analytical model for the coupling of OOPLs in TDI (see \cref{eq:ASDXeta}) depending on the laser noise level and the adjacent OOPLs in the RFI. This model can be of further importance for the study of underperforming lasers in LISA and for the assessment of laser requirements in next-generation space-based gravitational-wave missions. We validate the model numerically: We include OOPLs in the LISA simulation \cite{Bayle:LisaInstrument} and compute the associated laser noise residual in the second-generation TDI Michelson variable $X_2$ using PyTDI \cite{Staab:PyTDI}. The numerical results agree with the model.\par
We derive a compensation scheme for OOPLs (OOPL-CS), which includes OOPL delay and advancement operators in the TDI combinations. As a byproduct of the OOPL-CS, we derive a guideline for the OB design (see \cref{eq:OpticalBenchDesignGuideline}): To facilitate the complete cancellation of the OOPL-related laser noise residuals in TDI, the OOPL differences have to match between TMI and RFI. Any deviation from this design guideline will cause a residual laser noise proportional to the deviation. The current OB design values \cite{Brzozowski:LISA-OB} involve a mismatch of about \SI{2}{\centi\m}. We derive an analytical model for the impact of this mismatch and verify it numerically. It is in the order of \SI{10}{\femto\m \hertz\tothe{-0.5}} and thus negligible across the LISA band. This model can be of further importance for the OB design in next-generation space-based gravitational-wave missions, where it can be used to formulate concrete requirements.\par
We assess the performance of the OOPL-CS numerically. In a simulation with just laser frequency noise, we show that the OOPL-CS can completely remove the OOPL-related laser noise residuals. Furthermore, we validate the performance of the OOPL-CS in the presence of secondary noises and for underperforming lasers with increased laser frequency noise of \num{300} and \SI{3000}{\hertz \hertz\tothe{-0.5}}. In the realistic case of \SI{30}{\hertz \hertz\tothe{-0.5}}, the OOPL-related laser noise residuals reach a few \si{\pico\m \hertz\tothe{-0.5}} but remain below the secondary noises. Here OOPLs do not affect the performance. Nonetheless, every noise above \SI{1}{\pico \m \hertz\tothe{-0.5}} demands a careful evaluation, and it is recommended to explore the application of the OOPL-CS to ensure optimum performance. In the case of underperforming lasers, the OOPL-related laser noise residuals can surpass the secondary noises. Our results confirm that the OOPL-CS can completely remove the OOPL-related laser noise residuals in this scenario.\par
In our simulations, we consider six free-running lasers. In reality, the lasers are locked to each other~\cite{Heinzel:2024FrequencyPlanning}. We discuss the impact of different laser-locking configurations considering our analytical models. From the perspective of OOPL-related laser noise residuals, we identify suitable and unsuitable locking configurations. It would be interesting to numerically assess the impact of OOPLs with locked lasers in a follow-up investigation.\par
Furthermore, the OOPL-CS should be embedded in a full end-to-end pipeline, including all the effects we ignored (intersatellite ranging processing~\cite{Reinhardt:RangingSensorFusion}, clock desynchronizations~\cite{Reinhardt2024:ClockSynchronization}, clock-related noise~\cite{Hartwig:2021Clock}, tilt-to-length couplings~\cite{Paczkowski:2022TTL}, etc.).

\appendix
\section*{Acknowledgements}
The authors thank the LISA simulation expert group for all simulation-related activities, particularly for the development of LISA Instrument, LISA Orbits, and PyTDI. The authors thank Jean-Baptiste Bayle and Walter Fichter for useful discussions. The authors acknowledge support by the German Aerospace Center (DLR) with funds from the Federal Ministry of Economics and Technology (BMWi) according to a decision of the German Federal Parliament (Grant No. 50OQ2301, based on Grants No. 50OQ0601, No. 50OQ1301, No. 50OQ1801). Furthermore, this work was supported by the LEGACY cooperation on low-frequency gravitational wave astronomy (M.IF.A.QOP18098) and by the Bundesministerium für Wirtschaft und Klimaschutz based on a resolution of the German Bundestag (Project Ref. Number 50 OQ 1801). Jan Niklas Reinhardt acknowledges the funding by the Deutsche Forschungsgemeinschaft (DFG, German Research Foundation) under Germany's Excellence Strategy within the Cluster of Excellence PhoenixD (EXC 2122, Project ID 390833453). He also acknowledges the support of the IMPRS on Gravitational Wave Astronomy at the Max-Planck-Institut für Gravitationsphysik in Hannover, Germany. Philipp Euringer and Gerald Hechenblaikner acknowledge the funding from the Max-Planck-Institut für Gravitationsphysik (Albert-Einstein-Institut), based on a grant by the Deutsches Zentrum für Luft- und Raumfahrt (DLR). Kohei Yamamoto acknowledges support from the Cluster of Excellence “QuantumFrontiers: Light and Matter at the Quantum Frontier: Foundations and Applications in Metrology” (EXC-2123, Project No. 390837967).

\section{Manufacturing asymmetries}\label{app:ManufacturingAsymmetries}
\begin{figure}
	\begin{center}
		\includegraphics[width=0.5\textwidth]{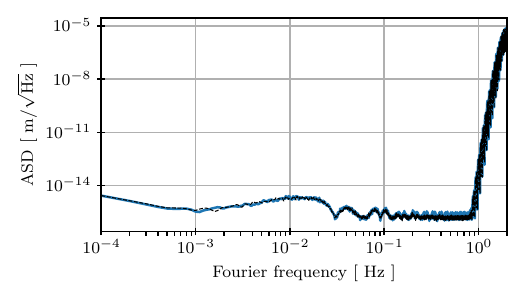}
	\end{center}
	\caption{Impact of manufacturing asymmetries: The blue and black plots show the ASDs of $X_2$ for simulations with and without manufacturing asymmetries in the order of \SI{100}{\micro\m}.}\label{fig:ImpactOfManufacturingAsymmetriesInTDI}
\end{figure}
This section investigates the laser noise residuals caused by manufacturing asymmetries between the OBs. We perform two simulations according to the setup with just laser frequency noise (see \cref{sec:ResultsJustLaserNoise}). In the first simulation, we set all OOPLs to zero. In the second one, we consider manufacturing asymmetries between the OBs: We draw all OOPLs from a zero-mean Gaussian distribution with \SI{100}{\micro\m} as standard deviation. For both simulations, we compute the second-generation TDI X Michelson combinations. Fig. \ref{fig:ImpactOfManufacturingAsymmetriesInTDI} shows the corresponding ASDs in \si{\m \hertz \tothe{-0.5}}. The impact of manufacturing asymmetries is in the order of \SI{1}{\femto\m \hertz \tothe{-0.5}}. It is completely negligible across the LISA band. Hence, OOPLs are well determined by their design values (first column \cref{tab:OnboardOpticalPathLengths}).

\section{Laser Locking}\label{app:LaserLocking}
The six LISA lasers are not independently free-running but offset frequency locked to each other according to a frequency plan \cite{Heinzel:2024FrequencyPlanning}. \cite{Bayle:LISA-Simulation} lists the six laser-locking configurations from the perspective of laser 12 as the primary laser. Our analytical models for the coupling of OOPLs in TDI suggest that certain laser-locking configurations naturally reduce the OOPL-related laser noise residuals.\par
We revisit the laser noise residuals in $X_2$ associated with uncompensated OOPLs in TDI step 2 (see \cref{eq:XEta}):
\begin{align}
    X^{\eta} &= \frac{d_{\text{rfi}}^{\text{adj}}}{2} \left(1 - 2\textbf{D}^2 + 2\textbf{D}^6 - \textbf{D}^8\right) \left(\ddot{p}_{12} - \ddot{p}_{13}\right),
\end{align}
$Y^{\eta}$ and $Z^{\eta}$ can be obtained via cyclic permutation of the SC indices. In these laser noise residuals, local and adjacent lasers appear with opposite signs. They, therefore, cancel each other in laser-locking configurations that lock local and adjacent lasers onto each other, assuming sufficient gain of the locking control loop. Thus, the laser-locking configurations N1-12, N3-12, and N5-12 naturally reduce the OOPL-related laser noise residuals $X^{\eta}$, $Y^{\eta}$, and $Z^{\eta}$ in the second-generation TDI Michelson variables. The locking schemes (N2-12, N4-12, N6-12), on the other hand, involve one pair of not directly locked local and adjacent lasers. Here, the OOPL-related laser noise residuals do not cancel in all three TDI channels.\par
We also revisit the laser noise residuals in $X_2$ associated with uncompensated OOPLs in TDI step 1 (see \cref{eq:TdiXxiCompleteResidual}):
\begin{align}
    &X^{\xi} = (1 - \textbf{D}^2 - \textbf{D}^4 + \textbf{D}^6)\nonumber\\
    &\Bigg{\{} \frac{d_{\Delta}^{\text{loc}} + d_{\Delta}^{\text{adj}}}{2} \left(1 + \textbf{D}^2\right) \left(\ddot{p}_{13} - \ddot{p}_{12}\right)\nonumber\\
    &+ d_{\Delta}^{\text{loc}} \: \textbf{D} \left(\ddot{p}_{31} - \ddot{p}_{21}\right)\nonumber\\ &+ d_{\Delta}^{\text{adj}} \: \textbf{D} \left(\ddot{p}_{23} - \ddot{p}_{32}\right) \Bigg{\}},
\end{align}
$Y^{\xi}$ and $Z^{\xi}$ can be obtained via cyclic permutation of the SC indices. In the first term, local and adjacent lasers appear with opposite signs. Hence, the laser-locking configurations N1-12, N3-12, and N5-12 naturally suppress its contribution. The remaining terms involve a pair of lasers on different spacecraft, which can only be locked onto each other with an inter-satellite delay. These terms, therefore, cannot be meaningfully suppressed and contribute to laser noise residuals in each laser-locking configuration.

\clearpage
\bibliography{references}

\begin{thebibliography}{28}%
\makeatletter
\providecommand \@ifxundefined [1]{%
 \@ifx{#1\undefined}
}%
\providecommand \@ifnum [1]{%
 \ifnum #1\expandafter \@firstoftwo
 \else \expandafter \@secondoftwo
 \fi
}%
\providecommand \@ifx [1]{%
 \ifx #1\expandafter \@firstoftwo
 \else \expandafter \@secondoftwo
 \fi
}%
\providecommand \natexlab [1]{#1}%
\providecommand \enquote  [1]{``#1''}%
\providecommand \bibnamefont  [1]{#1}%
\providecommand \bibfnamefont [1]{#1}%
\providecommand \citenamefont [1]{#1}%
\providecommand \href@noop [0]{\@secondoftwo}%
\providecommand \href [0]{\begingroup \@sanitize@url \@href}%
\providecommand \@href[1]{\@@startlink{#1}\@@href}%
\providecommand \@@href[1]{\endgroup#1\@@endlink}%
\providecommand \@sanitize@url [0]{\catcode `\\12\catcode `\$12\catcode
  `\&12\catcode `\#12\catcode `\^12\catcode `\_12\catcode `\%12\relax}%
\providecommand \@@startlink[1]{}%
\providecommand \@@endlink[0]{}%
\providecommand \url  [0]{\begingroup\@sanitize@url \@url }%
\providecommand \@url [1]{\endgroup\@href {#1}{\urlprefix }}%
\providecommand \urlprefix  [0]{URL }%
\providecommand \Eprint [0]{\href }%
\providecommand \doibase [0]{https://doi.org/}%
\providecommand \selectlanguage [0]{\@gobble}%
\providecommand \bibinfo  [0]{\@secondoftwo}%
\providecommand \bibfield  [0]{\@secondoftwo}%
\providecommand \translation [1]{[#1]}%
\providecommand \BibitemOpen [0]{}%
\providecommand \bibitemStop [0]{}%
\providecommand \bibitemNoStop [0]{.\EOS\space}%
\providecommand \EOS [0]{\spacefactor3000\relax}%
\providecommand \BibitemShut  [1]{\csname bibitem#1\endcsname}%
\let\auto@bib@innerbib\@empty
\bibitem [{\citenamefont {Colpi}\ \emph {et~al.}(2024)\citenamefont {Colpi}
  \emph {et~al.}}]{Colpi:2024RedBook}%
  \BibitemOpen
  \bibfield  {author} {\bibinfo {author} {\bibfnamefont {M.}~\bibnamefont
  {Colpi}} \emph {et~al.},\ }\bibfield  {title} {\bibinfo {title} {{LISA
  Definition Study Report}},\ }\href@noop {} {\  (\bibinfo {year} {2024})},\
  \Eprint {https://arxiv.org/abs/2402.07571} {arXiv:2402.07571 [astro-ph.CO]}
  \BibitemShut {NoStop}%
\bibitem [{\citenamefont {Heinzel}\ \emph {et~al.}(2024)\citenamefont
  {Heinzel}, \citenamefont {\'Alvarez-Vizoso}, \citenamefont
  {Dovale-\'Alvarez},\ and\ \citenamefont
  {Wiesner}}]{Heinzel:2024FrequencyPlanning}%
  \BibitemOpen
  \bibfield  {author} {\bibinfo {author} {\bibfnamefont {G.}~\bibnamefont
  {Heinzel}}, \bibinfo {author} {\bibfnamefont {J.}~\bibnamefont
  {\'Alvarez-Vizoso}}, \bibinfo {author} {\bibfnamefont {M.}~\bibnamefont
  {Dovale-\'Alvarez}},\ and\ \bibinfo {author} {\bibfnamefont {K.}~\bibnamefont
  {Wiesner}},\ }\bibfield  {title} {\bibinfo {title} {{Frequency planning for
  LISA}},\ }\href {https://doi.org/10.1103/PhysRevD.110.042002} {\bibfield
  {journal} {\bibinfo  {journal} {Phys. Rev. D}\ }\textbf {\bibinfo {volume}
  {110}},\ \bibinfo {pages} {042002} (\bibinfo {year} {2024})},\ \Eprint
  {https://arxiv.org/abs/2407.10621} {arXiv:2407.10621 [physics.ins-det]}
  \BibitemShut {NoStop}%
\bibitem [{\citenamefont {Gerberding}\ \emph {et~al.}(2013)\citenamefont
  {Gerberding}, \citenamefont {Sheard}, \citenamefont {Bykov}, \citenamefont
  {Kullmann}, \citenamefont {Delgado}, \citenamefont {Danzmann},\ and\
  \citenamefont {Heinzel}}]{Gerberding:Phasemeter}%
  \BibitemOpen
  \bibfield  {author} {\bibinfo {author} {\bibfnamefont {O.}~\bibnamefont
  {Gerberding}}, \bibinfo {author} {\bibfnamefont {B.}~\bibnamefont {Sheard}},
  \bibinfo {author} {\bibfnamefont {I.}~\bibnamefont {Bykov}}, \bibinfo
  {author} {\bibfnamefont {J.}~\bibnamefont {Kullmann}}, \bibinfo {author}
  {\bibfnamefont {J.~J.~E.}\ \bibnamefont {Delgado}}, \bibinfo {author}
  {\bibfnamefont {K.}~\bibnamefont {Danzmann}},\ and\ \bibinfo {author}
  {\bibfnamefont {G.}~\bibnamefont {Heinzel}},\ }\bibfield  {title} {\bibinfo
  {title} {Phasemeter core for intersatellite laser heterodyne interferometry:
  modelling, simulations and experiments},\ }\href@noop {} {\bibfield
  {journal} {\bibinfo  {journal} {Classical and Quantum Gravity}\ }\textbf
  {\bibinfo {volume} {30}},\ \bibinfo {pages} {235029} (\bibinfo {year}
  {2013})}\BibitemShut {NoStop}%
\bibitem [{\citenamefont {Armstrong}\ \emph {et~al.}(1999)\citenamefont
  {Armstrong}, \citenamefont {Estabrook},\ and\ \citenamefont
  {Tinto}}]{Armstrong:TDI}%
  \BibitemOpen
  \bibfield  {author} {\bibinfo {author} {\bibfnamefont {J.}~\bibnamefont
  {Armstrong}}, \bibinfo {author} {\bibfnamefont {F.}~\bibnamefont
  {Estabrook}},\ and\ \bibinfo {author} {\bibfnamefont {M.}~\bibnamefont
  {Tinto}},\ }\bibfield  {title} {\bibinfo {title} {Time-delay interferometry
  for space-based gravitational wave searches},\ }\href@noop {} {\bibfield
  {journal} {\bibinfo  {journal} {The Astrophysical Journal}\ }\textbf
  {\bibinfo {volume} {527}},\ \bibinfo {pages} {814} (\bibinfo {year}
  {1999})}\BibitemShut {NoStop}%
\bibitem [{\citenamefont {Tinto}\ \emph {et~al.}(2002)\citenamefont {Tinto},
  \citenamefont {Estabrook},\ and\ \citenamefont
  {Armstrong}}]{Tinto:TDIforLISA}%
  \BibitemOpen
  \bibfield  {author} {\bibinfo {author} {\bibfnamefont {M.}~\bibnamefont
  {Tinto}}, \bibinfo {author} {\bibfnamefont {F.~B.}\ \bibnamefont
  {Estabrook}},\ and\ \bibinfo {author} {\bibfnamefont {J.}~\bibnamefont
  {Armstrong}},\ }\bibfield  {title} {\bibinfo {title} {Time-delay
  interferometry for {LISA}},\ }\href@noop {} {\bibfield  {journal} {\bibinfo
  {journal} {Physical Review D}\ }\textbf {\bibinfo {volume} {65}},\ \bibinfo
  {pages} {082003} (\bibinfo {year} {2002})}\BibitemShut {NoStop}%
\bibitem [{\citenamefont {Tinto}\ and\ \citenamefont
  {Dhurandhar}(2021)}]{Tinto:2020TDIReview}%
  \BibitemOpen
  \bibfield  {author} {\bibinfo {author} {\bibfnamefont {M.}~\bibnamefont
  {Tinto}}\ and\ \bibinfo {author} {\bibfnamefont {S.~V.}\ \bibnamefont
  {Dhurandhar}},\ }\bibfield  {title} {\bibinfo {title} {{Time-delay
  interferometry}},\ }\href {https://doi.org/10.1007/s41114-020-00029-6}
  {\bibfield  {journal} {\bibinfo  {journal} {Living Rev. Rel.}\ }\textbf
  {\bibinfo {volume} {24}},\ \bibinfo {pages} {1} (\bibinfo {year}
  {2021})}\BibitemShut {NoStop}%
\bibitem [{\citenamefont {Heinzel}\ \emph {et~al.}(2011)\citenamefont
  {Heinzel}, \citenamefont {Esteban}, \citenamefont {Barke}, \citenamefont
  {Otto}, \citenamefont {Wang}, \citenamefont {Garcia},\ and\ \citenamefont
  {Danzmann}}]{Heinzel:Ranging}%
  \BibitemOpen
  \bibfield  {author} {\bibinfo {author} {\bibfnamefont {G.}~\bibnamefont
  {Heinzel}}, \bibinfo {author} {\bibfnamefont {J.~J.}\ \bibnamefont
  {Esteban}}, \bibinfo {author} {\bibfnamefont {S.}~\bibnamefont {Barke}},
  \bibinfo {author} {\bibfnamefont {M.}~\bibnamefont {Otto}}, \bibinfo {author}
  {\bibfnamefont {Y.}~\bibnamefont {Wang}}, \bibinfo {author} {\bibfnamefont
  {A.~F.}\ \bibnamefont {Garcia}},\ and\ \bibinfo {author} {\bibfnamefont
  {K.}~\bibnamefont {Danzmann}},\ }\bibfield  {title} {\bibinfo {title}
  {Auxiliary functions of the {LISA} laser link: ranging, clock noise transfer
  and data communication},\ }\href@noop {} {\bibfield  {journal} {\bibinfo
  {journal} {Classical and Quantum Gravity}\ }\textbf {\bibinfo {volume}
  {28}},\ \bibinfo {pages} {094008} (\bibinfo {year} {2011})}\BibitemShut
  {NoStop}%
\bibitem [{\citenamefont {Sutton}\ \emph {et~al.}(2010)\citenamefont {Sutton},
  \citenamefont {McKenzie}, \citenamefont {Ware},\ and\ \citenamefont
  {Shaddock}}]{Sutton:Ranging}%
  \BibitemOpen
  \bibfield  {author} {\bibinfo {author} {\bibfnamefont {A.}~\bibnamefont
  {Sutton}}, \bibinfo {author} {\bibfnamefont {K.}~\bibnamefont {McKenzie}},
  \bibinfo {author} {\bibfnamefont {B.}~\bibnamefont {Ware}},\ and\ \bibinfo
  {author} {\bibfnamefont {D.~A.}\ \bibnamefont {Shaddock}},\ }\bibfield
  {title} {\bibinfo {title} {Laser ranging and communications for lisa},\
  }\href {https://doi.org/10.1364/OE.18.020759} {\bibfield  {journal} {\bibinfo
   {journal} {Opt. Express}\ }\textbf {\bibinfo {volume} {18}},\ \bibinfo
  {pages} {20759} (\bibinfo {year} {2010})}\BibitemShut {NoStop}%
\bibitem [{\citenamefont {Armano}\ \emph {et~al.}(2016)\citenamefont {Armano},
  \citenamefont {Audley}, \citenamefont {Auger}, \citenamefont {Baird},
  \citenamefont {Bassan}, \citenamefont {Binetruy}, \citenamefont {Born},
  \citenamefont {Bortoluzzi}, \citenamefont {Brandt}, \citenamefont {Caleno}
  \emph {et~al.}}]{Armano:LPF-TestMasses}%
  \BibitemOpen
  \bibfield  {author} {\bibinfo {author} {\bibfnamefont {M.}~\bibnamefont
  {Armano}}, \bibinfo {author} {\bibfnamefont {H.}~\bibnamefont {Audley}},
  \bibinfo {author} {\bibfnamefont {G.}~\bibnamefont {Auger}}, \bibinfo
  {author} {\bibfnamefont {J.~T.}\ \bibnamefont {Baird}}, \bibinfo {author}
  {\bibfnamefont {M.}~\bibnamefont {Bassan}}, \bibinfo {author} {\bibfnamefont
  {P.}~\bibnamefont {Binetruy}}, \bibinfo {author} {\bibfnamefont
  {M.}~\bibnamefont {Born}}, \bibinfo {author} {\bibfnamefont {D.}~\bibnamefont
  {Bortoluzzi}}, \bibinfo {author} {\bibfnamefont {N.}~\bibnamefont {Brandt}},
  \bibinfo {author} {\bibfnamefont {M.}~\bibnamefont {Caleno}}, \emph
  {et~al.},\ }\bibfield  {title} {\bibinfo {title} {Sub-femto-g free fall for
  space-based gravitational wave observatories: {LISA} pathfinder results},\
  }\href@noop {} {\bibfield  {journal} {\bibinfo  {journal} {Physical review
  letters}\ }\textbf {\bibinfo {volume} {116}},\ \bibinfo {pages} {231101}
  (\bibinfo {year} {2016})}\BibitemShut {NoStop}%
\bibitem [{\citenamefont {Armano}\ \emph {et~al.}(2018)\citenamefont {Armano},
  \citenamefont {Audley}, \citenamefont {Baird}, \citenamefont {Binetruy},
  \citenamefont {Born}, \citenamefont {Bortoluzzi}, \citenamefont {Castelli},
  \citenamefont {Cavalleri}, \citenamefont {Cesarini}, \citenamefont {Cruise}
  \emph {et~al.}}]{Armano:BeyondLPF}%
  \BibitemOpen
  \bibfield  {author} {\bibinfo {author} {\bibfnamefont {M.}~\bibnamefont
  {Armano}}, \bibinfo {author} {\bibfnamefont {H.}~\bibnamefont {Audley}},
  \bibinfo {author} {\bibfnamefont {J.}~\bibnamefont {Baird}}, \bibinfo
  {author} {\bibfnamefont {P.}~\bibnamefont {Binetruy}}, \bibinfo {author}
  {\bibfnamefont {M.}~\bibnamefont {Born}}, \bibinfo {author} {\bibfnamefont
  {D.}~\bibnamefont {Bortoluzzi}}, \bibinfo {author} {\bibfnamefont
  {E.}~\bibnamefont {Castelli}}, \bibinfo {author} {\bibfnamefont
  {A.}~\bibnamefont {Cavalleri}}, \bibinfo {author} {\bibfnamefont
  {A.}~\bibnamefont {Cesarini}}, \bibinfo {author} {\bibfnamefont
  {A.}~\bibnamefont {Cruise}}, \emph {et~al.},\ }\bibfield  {title} {\bibinfo
  {title} {Beyond the required {LISA} free-fall performance: new {LISA}
  pathfinder results down to 20 $\mu$ hz},\ }\href@noop {} {\bibfield
  {journal} {\bibinfo  {journal} {Physical review letters}\ }\textbf {\bibinfo
  {volume} {120}},\ \bibinfo {pages} {061101} (\bibinfo {year}
  {2018})}\BibitemShut {NoStop}%
\bibitem [{\citenamefont {Otto}(2015)}]{Otto:Thesis}%
  \BibitemOpen
  \bibfield  {author} {\bibinfo {author} {\bibfnamefont {M.}~\bibnamefont
  {Otto}},\ }\bibfield  {title} {\bibinfo {title} {Time-delay interferometry
  simulations for the laser interferometer space antenna},\ }\href@noop {} {\
  (\bibinfo {year} {2015})}\BibitemShut {NoStop}%
\bibitem [{\citenamefont {Brzozowski}\ \emph {et~al.}(2022)\citenamefont
  {Brzozowski}, \citenamefont {Robertson}, \citenamefont {Fitzsimons},
  \citenamefont {Ward}, \citenamefont {Keogh}, \citenamefont {Taylor},
  \citenamefont {Milanova}, \citenamefont {Perreur-Lloyd}, \citenamefont {Ali},
  \citenamefont {Earle} \emph {et~al.}}]{Brzozowski:LISA-OB}%
  \BibitemOpen
  \bibfield  {author} {\bibinfo {author} {\bibfnamefont {W.}~\bibnamefont
  {Brzozowski}}, \bibinfo {author} {\bibfnamefont {D.}~\bibnamefont
  {Robertson}}, \bibinfo {author} {\bibfnamefont {E.}~\bibnamefont
  {Fitzsimons}}, \bibinfo {author} {\bibfnamefont {H.}~\bibnamefont {Ward}},
  \bibinfo {author} {\bibfnamefont {J.}~\bibnamefont {Keogh}}, \bibinfo
  {author} {\bibfnamefont {A.}~\bibnamefont {Taylor}}, \bibinfo {author}
  {\bibfnamefont {M.}~\bibnamefont {Milanova}}, \bibinfo {author}
  {\bibfnamefont {M.}~\bibnamefont {Perreur-Lloyd}}, \bibinfo {author}
  {\bibfnamefont {Z.}~\bibnamefont {Ali}}, \bibinfo {author} {\bibfnamefont
  {A.}~\bibnamefont {Earle}}, \emph {et~al.},\ }\bibfield  {title} {\bibinfo
  {title} {The {LISA} optical bench: an overview and engineering challenges},\
  }\href@noop {} {\bibfield  {journal} {\bibinfo  {journal} {Space Telescopes
  and Instrumentation 2022: Optical, Infrared, and Millimeter Wave}\ }\textbf
  {\bibinfo {volume} {12180}},\ \bibinfo {pages} {211} (\bibinfo {year}
  {2022})}\BibitemShut {NoStop}%
\bibitem [{\citenamefont {Euringer}\ \emph {et~al.}(2024)\citenamefont
  {Euringer}, \citenamefont {Houba}, \citenamefont {Hechenblaikner},
  \citenamefont {Mandel}, \citenamefont {Soualle},\ and\ \citenamefont
  {Fichter}}]{Euringer:FrontendAndModulationDelays}%
  \BibitemOpen
  \bibfield  {author} {\bibinfo {author} {\bibfnamefont {P.}~\bibnamefont
  {Euringer}}, \bibinfo {author} {\bibfnamefont {N.}~\bibnamefont {Houba}},
  \bibinfo {author} {\bibfnamefont {G.}~\bibnamefont {Hechenblaikner}},
  \bibinfo {author} {\bibfnamefont {O.}~\bibnamefont {Mandel}}, \bibinfo
  {author} {\bibfnamefont {F.}~\bibnamefont {Soualle}},\ and\ \bibinfo {author}
  {\bibfnamefont {W.}~\bibnamefont {Fichter}},\ }\bibfield  {title} {\bibinfo
  {title} {Compensation of front-end and modulation delays in phase and ranging
  measurements for time-delay interferometry},\ }\href@noop {} {\bibfield
  {journal} {\bibinfo  {journal} {Physical Review D}\ }\textbf {\bibinfo
  {volume} {109}},\ \bibinfo {pages} {083024} (\bibinfo {year}
  {2024})}\BibitemShut {NoStop}%
\bibitem [{\citenamefont {Yamamoto}\ \emph {et~al.}(2024)\citenamefont
  {Yamamoto}, \citenamefont {Bykov}, \citenamefont {Reinhardt}, \citenamefont
  {Bode}, \citenamefont {Grafe}, \citenamefont {Staab}, \citenamefont
  {Messied}, \citenamefont {Clark}, \citenamefont {Barranco}, \citenamefont
  {Schwarze} \emph {et~al.}}]{Yamamoto:2024Hexagon}%
  \BibitemOpen
  \bibfield  {author} {\bibinfo {author} {\bibfnamefont {K.}~\bibnamefont
  {Yamamoto}}, \bibinfo {author} {\bibfnamefont {I.}~\bibnamefont {Bykov}},
  \bibinfo {author} {\bibfnamefont {J.~N.}\ \bibnamefont {Reinhardt}}, \bibinfo
  {author} {\bibfnamefont {C.}~\bibnamefont {Bode}}, \bibinfo {author}
  {\bibfnamefont {P.}~\bibnamefont {Grafe}}, \bibinfo {author} {\bibfnamefont
  {M.}~\bibnamefont {Staab}}, \bibinfo {author} {\bibfnamefont
  {N.}~\bibnamefont {Messied}}, \bibinfo {author} {\bibfnamefont
  {M.}~\bibnamefont {Clark}}, \bibinfo {author} {\bibfnamefont {G.~F.}\
  \bibnamefont {Barranco}}, \bibinfo {author} {\bibfnamefont {T.~S.}\
  \bibnamefont {Schwarze}}, \emph {et~al.},\ }\bibfield  {title} {\bibinfo
  {title} {Experimental end-to-end demonstration of intersatellite absolute
  ranging for {LISA}},\ }\href@noop {} {\bibfield  {journal} {\bibinfo
  {journal} {arXiv preprint arXiv:2406.03074}\ } (\bibinfo {year}
  {2024})}\BibitemShut {NoStop}%
\bibitem [{\citenamefont {Reinhardt}\ \emph
  {et~al.}(2024{\natexlab{a}})\citenamefont {Reinhardt}, \citenamefont {Staab},
  \citenamefont {Yamamoto}, \citenamefont {Bayle}, \citenamefont {Hees},
  \citenamefont {Hartwig}, \citenamefont {Wiesner}, \citenamefont {Shah},\ and\
  \citenamefont {Heinzel}}]{Reinhardt:RangingSensorFusion}%
  \BibitemOpen
  \bibfield  {author} {\bibinfo {author} {\bibfnamefont {J.~N.}\ \bibnamefont
  {Reinhardt}}, \bibinfo {author} {\bibfnamefont {M.}~\bibnamefont {Staab}},
  \bibinfo {author} {\bibfnamefont {K.}~\bibnamefont {Yamamoto}}, \bibinfo
  {author} {\bibfnamefont {J.-B.}\ \bibnamefont {Bayle}}, \bibinfo {author}
  {\bibfnamefont {A.}~\bibnamefont {Hees}}, \bibinfo {author} {\bibfnamefont
  {O.}~\bibnamefont {Hartwig}}, \bibinfo {author} {\bibfnamefont
  {K.}~\bibnamefont {Wiesner}}, \bibinfo {author} {\bibfnamefont
  {S.}~\bibnamefont {Shah}},\ and\ \bibinfo {author} {\bibfnamefont
  {G.}~\bibnamefont {Heinzel}},\ }\bibfield  {title} {\bibinfo {title} {Ranging
  sensor fusion in {LISA} data processing: Treatment of ambiguities, noise, and
  onboard delays in {LISA} ranging observables},\ }\href
  {https://doi.org/10.1103/PhysRevD.109.022004} {\bibfield  {journal} {\bibinfo
   {journal} {Phys. Rev. D}\ }\textbf {\bibinfo {volume} {109}},\ \bibinfo
  {pages} {022004} (\bibinfo {year} {2024}{\natexlab{a}})}\BibitemShut
  {NoStop}%
\bibitem [{\citenamefont {Staab}\ \emph {et~al.}(2024)\citenamefont {Staab},
  \citenamefont {Lilley}, \citenamefont {Bayle},\ and\ \citenamefont
  {Hartwig}}]{Staab:LaserNoiseResiduals}%
  \BibitemOpen
  \bibfield  {author} {\bibinfo {author} {\bibfnamefont {M.}~\bibnamefont
  {Staab}}, \bibinfo {author} {\bibfnamefont {M.}~\bibnamefont {Lilley}},
  \bibinfo {author} {\bibfnamefont {J.-B.}\ \bibnamefont {Bayle}},\ and\
  \bibinfo {author} {\bibfnamefont {O.}~\bibnamefont {Hartwig}},\ }\bibfield
  {title} {\bibinfo {title} {Laser noise residuals in {LISA} from on-board
  processing and time-delay interferometry},\ }\href@noop {} {\bibfield
  {journal} {\bibinfo  {journal} {Physical Review D}\ }\textbf {\bibinfo
  {volume} {109}},\ \bibinfo {pages} {043040} (\bibinfo {year}
  {2024})}\BibitemShut {NoStop}%
\bibitem [{\citenamefont {Staab}(2023)}]{Staab:Thesis}%
  \BibitemOpen
  \bibfield  {author} {\bibinfo {author} {\bibfnamefont {M.~B.}\ \bibnamefont
  {Staab}},\ }\bibfield  {title} {\bibinfo {title} {Time-delay interferometric
  ranging for {LISA}: Statistical analysis of bias-free ranging using laser
  noise minimization},\ }\href@noop {} {\  (\bibinfo {year}
  {2023})}\BibitemShut {NoStop}%
\bibitem [{\citenamefont {Bayle}\ \emph {et~al.}(2021)\citenamefont {Bayle},
  \citenamefont {Hartwig},\ and\ \citenamefont
  {Staab}}]{Bayle:TDI-in-Frequency}%
  \BibitemOpen
  \bibfield  {author} {\bibinfo {author} {\bibfnamefont {J.-B.}\ \bibnamefont
  {Bayle}}, \bibinfo {author} {\bibfnamefont {O.}~\bibnamefont {Hartwig}},\
  and\ \bibinfo {author} {\bibfnamefont {M.}~\bibnamefont {Staab}},\ }\bibfield
   {title} {\bibinfo {title} {Adapting time-delay interferometry for {LISA}
  data in frequency},\ }\href@noop {} {\bibfield  {journal} {\bibinfo
  {journal} {Physical Review D}\ }\textbf {\bibinfo {volume} {104}},\ \bibinfo
  {pages} {023006} (\bibinfo {year} {2021})}\BibitemShut {NoStop}%
\bibitem [{\citenamefont {Reinhardt}\ \emph
  {et~al.}(2024{\natexlab{b}})\citenamefont {Reinhardt}, \citenamefont
  {Hartwig},\ and\ \citenamefont
  {Heinzel}}]{Reinhardt2024:ClockSynchronization}%
  \BibitemOpen
  \bibfield  {author} {\bibinfo {author} {\bibfnamefont {J.~N.}\ \bibnamefont
  {Reinhardt}}, \bibinfo {author} {\bibfnamefont {O.}~\bibnamefont {Hartwig}},\
  and\ \bibinfo {author} {\bibfnamefont {G.}~\bibnamefont {Heinzel}},\ }\href
  {https://arxiv.org/abs/2408.09832} {\bibinfo {title} {Clock synchronization
  and light-travel-time estimation for space-based gravitational-wave
  detectors}} (\bibinfo {year} {2024}{\natexlab{b}}),\ \Eprint
  {https://arxiv.org/abs/2408.09832} {arXiv:2408.09832 [gr-qc]} \BibitemShut
  {NoStop}%
\bibitem [{\citenamefont {Bayle}\ and\ \citenamefont
  {Hartwig}(2023)}]{Bayle:LISA-Simulation}%
  \BibitemOpen
  \bibfield  {author} {\bibinfo {author} {\bibfnamefont {J.-B.}\ \bibnamefont
  {Bayle}}\ and\ \bibinfo {author} {\bibfnamefont {O.}~\bibnamefont
  {Hartwig}},\ }\bibfield  {title} {\bibinfo {title} {Unified model for the
  {LISA} measurements and instrument simulations},\ }\href@noop {} {\bibfield
  {journal} {\bibinfo  {journal} {Physical Review D}\ }\textbf {\bibinfo
  {volume} {107}},\ \bibinfo {pages} {083019} (\bibinfo {year}
  {2023})}\BibitemShut {NoStop}%
\bibitem [{\citenamefont {Inchausp\'e}\ \emph {et~al.}(2022)\citenamefont
  {Inchausp\'e}, \citenamefont {Hewitson}, \citenamefont {Sauter},\ and\
  \citenamefont {Wass}}]{Inchauspe:2022SCjitter}%
  \BibitemOpen
  \bibfield  {author} {\bibinfo {author} {\bibfnamefont {H.}~\bibnamefont
  {Inchausp\'e}}, \bibinfo {author} {\bibfnamefont {M.}~\bibnamefont
  {Hewitson}}, \bibinfo {author} {\bibfnamefont {O.}~\bibnamefont {Sauter}},\
  and\ \bibinfo {author} {\bibfnamefont {P.}~\bibnamefont {Wass}},\ }\bibfield
  {title} {\bibinfo {title} {{New {LISA} dynamics feedback control scheme:
  Common-mode isolation of test mass control and probes of test-mass
  acceleration}},\ }\href {https://doi.org/10.1103/PhysRevD.106.022006}
  {\bibfield  {journal} {\bibinfo  {journal} {Phys. Rev. D}\ }\textbf {\bibinfo
  {volume} {106}},\ \bibinfo {pages} {022006} (\bibinfo {year} {2022})},\
  \Eprint {https://arxiv.org/abs/2202.12735} {arXiv:2202.12735 [gr-qc]}
  \BibitemShut {NoStop}%
\bibitem [{\citenamefont {Bayle}\ \emph
  {et~al.}(2022{\natexlab{a}})\citenamefont {Bayle}, \citenamefont {Hees},
  \citenamefont {Lilley}, \citenamefont {Le~Poncin-Lafitte}, \citenamefont
  {Martens},\ and\ \citenamefont {Joffre}}]{Bayle:LisaOrbits}%
  \BibitemOpen
  \bibfield  {author} {\bibinfo {author} {\bibfnamefont {J.-B.}\ \bibnamefont
  {Bayle}}, \bibinfo {author} {\bibfnamefont {A.}~\bibnamefont {Hees}},
  \bibinfo {author} {\bibfnamefont {M.}~\bibnamefont {Lilley}}, \bibinfo
  {author} {\bibfnamefont {C.}~\bibnamefont {Le~Poncin-Lafitte}}, \bibinfo
  {author} {\bibfnamefont {W.}~\bibnamefont {Martens}},\ and\ \bibinfo {author}
  {\bibfnamefont {E.}~\bibnamefont {Joffre}},\ }\href
  {https://doi.org/10.5281/zenodo.7700361} {\bibinfo {title} {L{ISA} {O}rbits}}
  (\bibinfo {year} {2022}{\natexlab{a}})\BibitemShut {NoStop}%
\bibitem [{\citenamefont {Paczkowski}\ \emph {et~al.}(2022)\citenamefont
  {Paczkowski}, \citenamefont {Giusteri}, \citenamefont {Hewitson},
  \citenamefont {Karnesis}, \citenamefont {Fitzsimons}, \citenamefont
  {Wanner},\ and\ \citenamefont {Heinzel}}]{Paczkowski:2022TTL}%
  \BibitemOpen
  \bibfield  {author} {\bibinfo {author} {\bibfnamefont {S.}~\bibnamefont
  {Paczkowski}}, \bibinfo {author} {\bibfnamefont {R.}~\bibnamefont
  {Giusteri}}, \bibinfo {author} {\bibfnamefont {M.}~\bibnamefont {Hewitson}},
  \bibinfo {author} {\bibfnamefont {N.}~\bibnamefont {Karnesis}}, \bibinfo
  {author} {\bibfnamefont {E.}~\bibnamefont {Fitzsimons}}, \bibinfo {author}
  {\bibfnamefont {G.}~\bibnamefont {Wanner}},\ and\ \bibinfo {author}
  {\bibfnamefont {G.}~\bibnamefont {Heinzel}},\ }\bibfield  {title} {\bibinfo
  {title} {Postprocessing subtraction of tilt-to-length noise in {LISA}},\
  }\href@noop {} {\bibfield  {journal} {\bibinfo  {journal} {Physical Review
  D}\ }\textbf {\bibinfo {volume} {106}},\ \bibinfo {pages} {042005} (\bibinfo
  {year} {2022})}\BibitemShut {NoStop}%
\bibitem [{\citenamefont {Hartwig}\ and\ \citenamefont
  {Bayle}(2021)}]{Hartwig:2021Clock}%
  \BibitemOpen
  \bibfield  {author} {\bibinfo {author} {\bibfnamefont {O.}~\bibnamefont
  {Hartwig}}\ and\ \bibinfo {author} {\bibfnamefont {J.-B.}\ \bibnamefont
  {Bayle}},\ }\bibfield  {title} {\bibinfo {title} {Clock-jitter reduction in
  {LISA} time-delay interferometry combinations},\ }\href@noop {} {\bibfield
  {journal} {\bibinfo  {journal} {Physical Review D}\ }\textbf {\bibinfo
  {volume} {103}},\ \bibinfo {pages} {123027} (\bibinfo {year}
  {2021})}\BibitemShut {NoStop}%
\bibitem [{\citenamefont {Risken}(1996)}]{Risken:PSD}%
  \BibitemOpen
  \bibfield  {author} {\bibinfo {author} {\bibfnamefont {H.}~\bibnamefont
  {Risken}},\ }\href@noop {} {\emph {\bibinfo {title} {Fokker-planck
  equation}}}\ (\bibinfo  {publisher} {Springer},\ \bibinfo {year}
  {1996})\BibitemShut {NoStop}%
\bibitem [{\citenamefont {Staab}\ \emph {et~al.}(2023)\citenamefont {Staab},
  \citenamefont {Bayle},\ and\ \citenamefont {Hartwig}}]{Staab:PyTDI}%
  \BibitemOpen
  \bibfield  {author} {\bibinfo {author} {\bibfnamefont {M.}~\bibnamefont
  {Staab}}, \bibinfo {author} {\bibfnamefont {J.-B.}\ \bibnamefont {Bayle}},\
  and\ \bibinfo {author} {\bibfnamefont {O.}~\bibnamefont {Hartwig}},\ }\href
  {https://doi.org/10.5281/zenodo.7704609} {\bibinfo {title} {Py{TDI}}}
  (\bibinfo {year} {2023})\BibitemShut {NoStop}%
\bibitem [{\citenamefont {Bayle}\ \emph
  {et~al.}(2022{\natexlab{b}})\citenamefont {Bayle}, \citenamefont {Hartwig},\
  and\ \citenamefont {Staab}}]{Bayle:LisaInstrument}%
  \BibitemOpen
  \bibfield  {author} {\bibinfo {author} {\bibfnamefont {J.-B.}\ \bibnamefont
  {Bayle}}, \bibinfo {author} {\bibfnamefont {O.}~\bibnamefont {Hartwig}},\
  and\ \bibinfo {author} {\bibfnamefont {M.}~\bibnamefont {Staab}},\ }\href
  {https://doi.org/10.5281/zenodo.7071251} {\bibinfo {title} {L{ISA}
  {I}nstrument}} (\bibinfo {year} {2022}{\natexlab{b}})\BibitemShut {NoStop}%
\bibitem [{\citenamefont {Martens}\ and\ \citenamefont
  {Joffre}(2021)}]{Martens:2021TrajectoryDesign}%
  \BibitemOpen
  \bibfield  {author} {\bibinfo {author} {\bibfnamefont {W.}~\bibnamefont
  {Martens}}\ and\ \bibinfo {author} {\bibfnamefont {E.}~\bibnamefont
  {Joffre}},\ }\bibfield  {title} {\bibinfo {title} {{Trajectory Design for the
  ESA LISA Mission}}\ }\href {https://doi.org/10.1007/s40295-021-00263-2}
  {10.1007/s40295-021-00263-2} (\bibinfo {year} {2021}),\ \Eprint
  {https://arxiv.org/abs/2101.03040} {arXiv:2101.03040 [gr-qc]} \BibitemShut
  {NoStop}%
\end{thebibliography}%
\end{document}